\documentclass[APA]{USG-editedBA}\usepackage{knitr} 
\articletype{Manuscript Submission}%
\received{00}
\revised{00}
\accepted{00}
\journal{}
\volume{0}
\copyyear{2026}
\startpage{1}
\articledoi{10.1002/0000}
\usepackage{colortbl}   
\usepackage[table]{xcolor}
\usepackage{array}
\usepackage{tikz}
\usepackage{amsmath,amssymb,amstext,amsthm}
\usepackage{wrapfig}
\usepackage{mathtools}
\usepackage{pgfplots}
\usepackage{float}
\usepackage{booktabs}
\usepackage{pdflscape}
\usepackage[aboveskip=1pt,labelfont=bf,labelsep=period,justification=raggedright,singlelinecheck=off]{caption}
\pgfplotsset{compat=1.18}
\usetikzlibrary{arrows.meta}
\usepgfplotslibrary{groupplots}
\usepgfplotslibrary{patchplots}
\graphicspath{ {./images/} }

\newcommand{\blandscape}{\begin{landscape}}
\newcommand{\elandscape}{\end{landscape}}
\IfFileExists{upquote.sty}{\usepackage{upquote}}{}
\begin{document}
\title{Explaining Human Choices with Simple Vector Representations}
\subtitle{An Investigation of Hiding and Seeking}
\author[1]{Peter A. V. DiBerardino}
\author[1,2]{Britt Anderson}
\authormark{DiBerardino \textsc{et al.}}
\titlemark{Explaining Human Choices with Simple Vector Representations}
\address[1]{\orgdiv{Dept. of Psychology,}\orgname{University of Waterloo,}\orgaddress{Waterloo, \state{ON }\country{Canada}}}
\address[2]{\orgdiv{Centre for Theoretical Neuroscience,}\orgname{University of Waterloo,}\orgaddress{Waterloo, \state{ON } \country{Canada}}}
\corres{Britt Anderson \email{britt@uwaterloo.ca}}
\fundingInfo{NSERC}
\abstract{We formalize human choice behavior in a probabilistic hide-and-seek task. In our geometric construction, vectors represent participant choice frequencies as well as probability matching and maximizing strategies. We measured choice behavior not just in the well-studied scenario of pursuing an objective (seeking), but also the rarely studied scenario of avoiding consequences (hiding). We used our geometric construction to define the avoidance counterpart of probability matching, \emph{probability antimatching}, as a vector reflection across the uniform distribution. Decomposing the behavior of participants when they were seeking into matching and maximizing components, we could mathematically derive the analogous antimatching and minimizing strategies for hiding. 

Participants did change their choice frequencies between hiding and seeking conditions. In both cases, we found that a linear combination of just two vectors did an excellent job of fitting participant choice frequencies: matching + maximizing for seeking, antimatching + minimizing for hiding. We could account for diversity in participant strategy usage by varying the coefficients of the two relevant basis strategy vectors. We successfully applied this model in scenarios of up to 7 rooms. We conclude that an apparent diversity of human conduct in stochastic environments can, in some cases, be explained by varying the weighting of two principle strategies: whether to match/antimatch or maximize/minimize.} 

\maketitle

\section{Introduction}

We implicitly learn to exploit the statistical
structure of our environments. We do this to build context-dependent conceptual structures (e.g. in language the transition probabilities between syllables are dependent on the language \cite{Saffran1996}). We do this to optimize the match between circumstance and action (this is the basis for reinforcement learning as a model for behavior \cite{Sutton2019}). It might be tempting to assume that the way we do this is by representing the underlying probability distributions directly. However, this account seems at odds with our computational and storage limitations, and all the probability fallacies that we are heir to (nicely reviewed by \cite{Huang_2024} in their introduction).

Despite these fallacies and limits we do seem to have the ability to track frequencies for modest numbers of options, suggesting some representation of the generating distribution. But this represention of the generating distribution cannot be the  sole determinant of behavior, because the same person behaves differently in different circumstances with same underlying distribution, and different people behave differently in the same circumstance with the same underlying distribution. We propose this behavioral variety within and between individuals can be explained by the generation of a small set (in fact just four) strategies that are derived, using simple Euclidean geometry, from the single learned generating distribution. 

To gather the data needed to test this geometric account, to be elaborated shortly, we had people play a game of hide-and-seek. It took place on a computer against a simulated child, e.g. Sally, that consistently hid and sought with a fixed distribution. Experimentally, each option was a room. The number of rooms varied by condition. The participant's role in the experiment could be hider or seeker, and by playing against different ``children'', the stochastic environment (distribution) could be varied. 

Most prior empirical work has focused on what would be for us, the seeking condition: participants make selections to obtain rewards. There, it is common to see probability matching \cite{vulkan2000economist}. Probability matching refers to selecting options in proportion to their occurrence in repeated choice experiments. As one concrete example, \citeA{Koehler2009} found a predominance of probability matching for a marble guessing game where participants were told that there were 30 green and 10 red marbles in a computerized urn task. Participants were given the motivation of 50 cents for each correct guess and asked to predict the outcome of subsequent trials. Allowing participants to observe the sequence of draws in order to search for patterns did not change their propensity to probability match.

Probability matching shows us a couple of things. First, people learn frequencies when there are modest number of options. Second, they are not optimal. In a tasks like these, the optimal strategy is maximizing: pick the option that is most frequent. For the \citeA{Koehler2009} experiment, exclusively picking the most likely option wins you 50 cents 75\% of the time ($1 * 0.75 + 0 * 0.25 = 75\%$) while probability matching only wins you 50 cents 62.5\% of the time ($0.75^2~+~0.25^2~=~62.5\%$). For studying probability representation though, this ``bug is a feature.''  If people always maximized we would know that they had learned the most probable option, but we would be blind to their representation of the other options. Observing probability matching, while not the optimal strategy, indicated that participants have learned the underlying probability distribution.

The prototypical choice pattern when people are given tasks like ``Where is the candy bar?'' or ``Which is the hot slot machine?'' is to choose with the same frequency as the observed events. But what if the scenario is slightly altered? What if instead of choosing the lever that releases food, the task is to avoid the lever that gives a shock? If instead of choosing the option that gives you a candy bar, your task is to avoid the goblet with the iocane powder? What would probability matching consist of then? You would not want to choose bad outcomes in proportion to their frequency of recurrence. Maybe you would \emph{antimatch}. In the two option scenario this is easy. If you see two options with a ratio of 3:1 where the more common one is the more noxious one, you pick it with a ratio of 1:3. Your choices are now the ``flipped'' version of the observed frequencies. Note, analogous to matching, antimatching is not the optimal strategy. Exclusively choosing the least likely option is. Call it minimizing. 

This reversal of the scenario would be interesting as a way to probe some of our conceptions about choice. As often framed in Bayesian decision processes \cite{alma9915625002181} the representation of probability is independent of the cost function. For a particular distribution of events we can change the gains and losses without changing the frequencies. Thus, simply due to symmetry considerations, we might expect in scenarios with a 3:1 ratio of outcomes to see choices switch from a matching pattern to antimatching pattern when we invert the gains to be losses, even if the representation of probabilities remain the same.

Of course, assuming that people invert gains and losses symmetrically is a strong assumption. The difference between failing to catch a prey and being caught by the predator are not equal in magnitude. If avoiding large losses, we might expect a \emph{probability minimizing} pattern to emerge. This is where the prey always chooses the option with the lowest probability, analogous to probability maximizing in pursuit contexts. Intuitively, many scenarios, such as hide-and-seek, seem amenable to such a predator-prey framing. 

These considerations have given us just a handful of frequency distributions to consider: matching/antimatching and maximizing/minimizing. Matching and maximizing are in play when we seek (predator), and antimatching and minimizing are in play when we hide (prey). From the literature, we know that people use the matching strategy, indicating they are sensitive to the task probabilities. The maximizing and minimizing strategies are easily determined from the presented distribution by taking the maximum or minimum probability option. We will present shortly how we might calculate the antimatching strategy. The antimatching strategy, maximizing and minimizing distributions can all be directly calculated from the matching probability distribution, which we already know is frequently utilized. 

These four strategies organize into two pairs when we consider our role in the scenario: seeking or hiding; predator or prey. But the observed behavior does not have to be one of these extremes. It can be a combination of them. The particular weighting of each strategy of the pair would be individually and situationally variable, and depend on numerous factors such as one's state of fatigue or distraction, the actual costs/losses, and perhaps also one's interest in information seeking. The ``predator'', for whom the costs of failure are generally low, may value learning what to expect by probing locations other than the most likely. Additionally, such probes may alert the predator to changes in the prey's strategy for choosing where to hide or simply a drift in strategy \cite{Koulouriotis_2008}. This distributional learning is revealed by probability matching. On the other side, in an environment where multiple rounds are the rule, a prey might show a strong bias for choosing less frequent predator choices, but still occassionally choose a popular predator spot as there is still a chance for information gain and change detection by refining the estimate for this option.

This idea is that the frequency of our choices in games like these, which superficially can seem highly variable, may be just the outcome of mixing two \emph{basis} strategies. The advantages of this idea are several. First, we don't need to debate whether we are or are not really Bayesian when we have a simple and flexible procedure on-hand for getting us most of the way there: a geometric model rather than a fully probabilistic one. Second, with the basis strategies easily determined from the experimental scenario we can derive a single statistic, the mixing coefficient,  for comparing the effect of variables like fatigue or cost on choice behavior. 

To implement this idea we need to formalize what is meant by an antimatching strategy, and how to calculate that from the matching strategy. The intuitions we may have from thinking about the two dimensional case (3:1 -> 1:3) are not honored in higher dimensions. There is no mathematically canonical``opposite'' of a probability distribution.

Our response is to define the antimatching strategy via a geometric framing and then test it by comparing to experimentally measured choice frequencies. The reason to rely on geometric constructions is partly because of their convenience: they are often easier for many of us to think about than abstract measure spaces, but also because vectors and geometry seem to be a fruitful tool for building neural and cognitive models  \cite{georgopoulos1989mental, gardenfors2004conceptual, eliasmith2013build, kriegeskorte2013representational, diberardino2024plinko, bauschke2022minimal, filipowicz2016adapting}.We present the geometric ideas here, in the introduction, because they are an important prelude to what we measured in our experiments and how we report the results. 

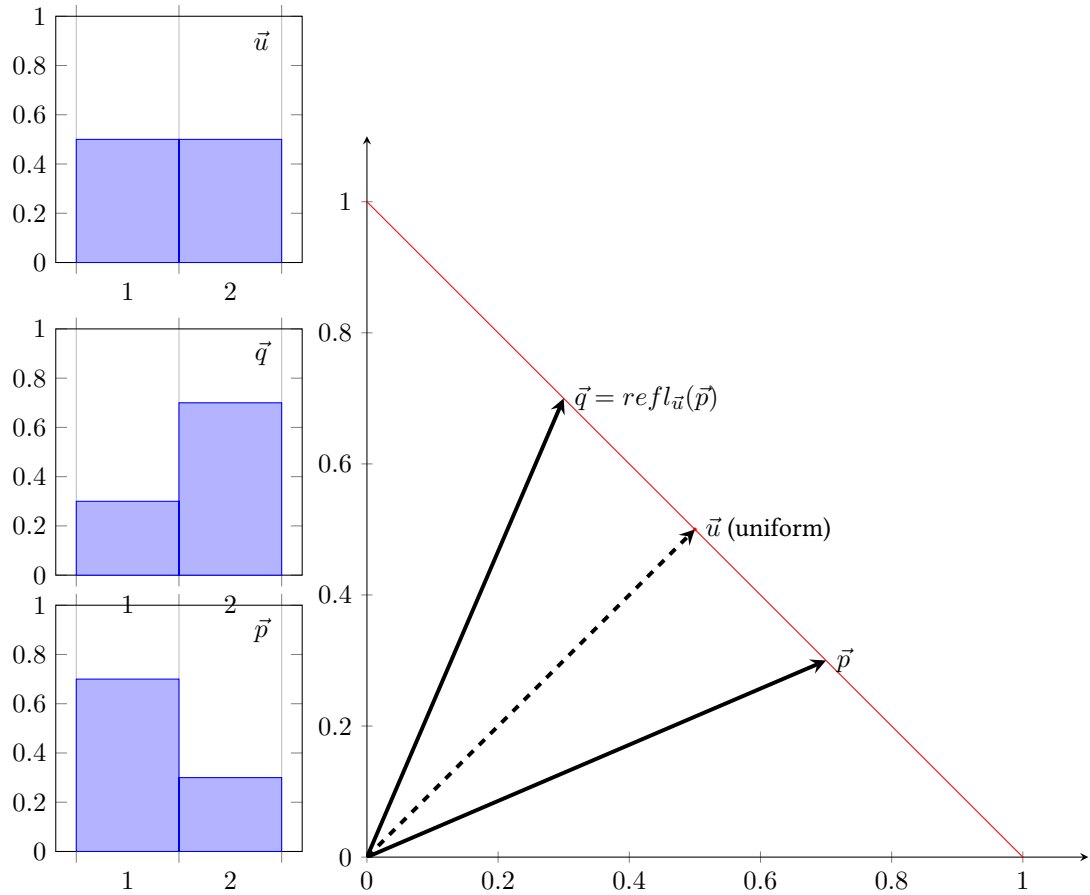
\begin{figure}[!htbp]
\centering
\subfloat{%
  \begin{tikzpicture}
    \begin{axis}[clip=false,ybar interval, ymax=1, ymin=0, width=4cm, height=4cm]
      \addplot coordinates {(1,0.30) (2,0.70) (3,0.20)};
      \node at (axis cs:2.8,0.9) {$\vec{q}$};
    \end{axis}
  \end{tikzpicture}%
}\hfill
\subfloat{%
  \begin{tikzpicture}
    \begin{axis}[clip=false,ybar interval, ymax=1, ymin=0, width=4cm, height=4cm]
      \addplot coordinates {(1,0.5) (2,0.50) (3,0.20)};
      \node at (axis cs:2.8,0.9) {$\vec{u}$};
    \end{axis}
  \end{tikzpicture}%
}\hfill
\subfloat{%
  \begin{tikzpicture}
    \begin{axis}[clip=false,ybar interval, ymax=1, ymin=0, width=4cm, height=4cm]
      \addplot coordinates {(1,0.70) (2,0.30) (3,0.20)};
      \node at (axis cs:2.8,0.9) {$\vec{p}$};
    \end{axis}
  \end{tikzpicture}%
}

\vspace{1ex}
\centering

\begin{tikzpicture}
  \begin{axis}[
      clip=false,
      width=0.7\textwidth,
      axis lines=middle,
      axis equal image,
      xmin=0, xmax=1.2,
      ymin=0, ymax=1.2,
      xtick=\empty, ytick=\empty,
    ]
    \addplot[domain=0:1, color=red, samples=2] {1-x};
    \addplot[mark=*, color=red, mark size=1pt] coordinates {(0.5,0.5)};

    \addplot[->, >=Stealth, line width=1.5pt, dashed]
      coordinates {(0,0) (0.5,0.5)}
      node[pos=1, anchor=west]{$\vec{u}$ (uniform)};
    \addplot[->, >=Stealth, line width=1.5pt]
      coordinates {(0,0) (0.7,0.3)}
      node[pos=1, anchor=west]{$\vec{p}$};
    \addplot[->, >=Stealth, line width=1.5pt]
      coordinates {(0,0) (0.3,0.7)}
      node[pos=1, anchor=west]{$\vec{q} = \mathrm{refl}_{\vec{u}}(\vec{p})$};
  \end{axis}
\end{tikzpicture}

\caption{The simplest case: a two-dimensional histogram.  On the top there are three histograms showing the proportion of events for each of two bins. Histograms (such as $\vec{p}$) can be reflected over the \emph{uniform}, denoted as $\vec{u}$, to obtain a $\vec{q}$, and vice versa. The angle between $\vec{q}$ and $\vec{u}$ is identical to the angle between $\vec{p}$ and $\vec{u}$. All of the vectors exist in the same plane. Two distributions are opposites if they are an equal angular distance away from the uniform vector and form a 2D-plane with the uniform.}
\label{fig:histo2dintro}
\end{figure}

First, consider the case of a two-dimensional game such as matching pennies (\url{https://en.wikipedia.org/wiki/Matching_pennies}; Figure \ref{fig:histo2dintro}).  The choices you make can be plotted as a two-dimensional histogram of the counts of your choices or the frequency with which you chose each option. This can be depicted as a vector on a two-dimensional plane. All permissible distributions live 
in the positive quadrant of 2D-space (you can never make a negative number of choices). The ``maximally uncertain'' distribution is the uniform, i.e. the histogram of choice frequencies where all the bars are of equal height. Its corresponding vector representation points to (0.5,0.5). For any given choice vector, we define the opposite choice vector to be the one that is on the opposite side of, and equidistant from, the maximally uncertain uniform vector. This is achieved by reflecting our choice vector across the uniform. The choice of the uniform vector as the reflection axis is in analogy to the uniform probability distribution in a Bayesian analysis.

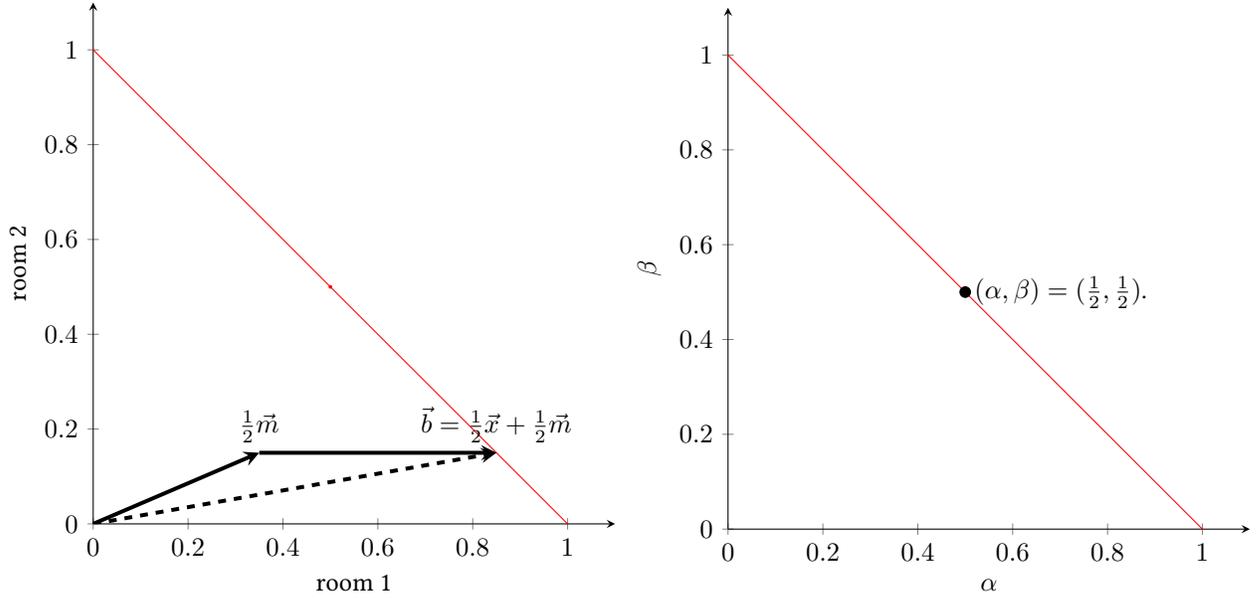
\begin{figure}[!htbp]
\centering
\subfloat[]{%
  \begin{tikzpicture}
    \begin{axis}[
        clip=false,
        width=0.45\textwidth,
        axis lines=middle,
        axis equal image,
        xmin=0, xmax=1.1,
        ymin=0, ymax=1.1,
        xtick=\empty, ytick=\empty,
        xlabel={room 1}, ylabel={room 2},
        xlabel style={anchor=west}, ylabel style={anchor=south},
      ]
      \addplot[domain=0:1, color=red, samples=2] {1-x};
      \addplot[mark=*, color=red, mark size=1pt] coordinates {(0.5,0.5)};

      \addplot[->, >=Stealth, line width=1.5pt, dashed]
        coordinates {(0,0) (0.85,0.15)}
        node[pos=1, anchor=south]{$\vec{b} = \tfrac{1}{2}\vec{x}+\tfrac{1}{2}\vec{m}$};
      \addplot[->, >=Stealth, line width=1.5pt]
        coordinates {(0.35,0.15) (0.85,0.15)};
      \addplot[->, >=Stealth, line width=1.5pt]
        coordinates {(0,0) (0.35,0.15)}
        node[pos=1, anchor=south]{$\tfrac{1}{2}\vec{m}$};
    \end{axis}
  \end{tikzpicture}%
}\hfill
\subfloat[]{%
  \begin{tikzpicture}
    \begin{axis}[
        clip=false,
        width=0.45\textwidth,
        axis lines=middle,
        axis equal image,
        xmin=0, xmax=1.1,
        ymin=0, ymax=1.1,
        xtick=\empty, ytick=\empty,
        xlabel={$\alpha$}, ylabel={$\beta$},
        xlabel style={anchor=west}, ylabel style={anchor=south},
      ]
      \addplot[domain=0:1, color=red, samples=2] {1-x};
      \addplot[mark=*, color=black, mark size=2pt]
        coordinates {(0.5,0.5)}
        node[anchor=west]{$(\alpha,\beta)=(\tfrac{1}{2},\tfrac{1}{2})$};
    \end{axis}
  \end{tikzpicture}%
}

\caption{From the space of histograms to the space of strategies: The counts of choices or their frequencies can be represented as a vector. This vector can be decomposed into the sum of the vector for the matching strategy and the maximizing strategy vector. On the left you see  $\vec{b} = (0.85, 0.15)$ and its decomposition of $\vec{b} = \alpha \vec{x} + \beta \vec{m}$ where $\vec{x} = (1,0)$ and $\vec{m} = (0.7, 0.3)$. The coefficients for each of these two strategies gives us a point on the  $(\alpha, \beta)$, in this case $ (\alpha, \beta) = (\frac{1}{2}, \frac{1}{2})$. Although the graph on the right looks similar to the one on the left \emph{the axes represent completely different things}. On the left we are in the space of options: one axis for each option. For a problem with seven choice options this graph would need to seven dimensions. Whereas on the right we are in the space of strategy. This plot will always only need two axes: one axis for each strategy: maximizing ($\alpha$) and matching ($\beta$). The same method works for decomposing hiding choices into a combination of antimatching and minimizing.}
\label{fig:stratspace}
\end{figure}

While it is easy to depict visually the vectors of two or three dimensional distributions, this becomes impossible as the number of options increases. In order to compare higher dimension histograms to each other and to compare conditions with varying numbers of options we would like to have a way to collapse performance to a consistent low dimensional representation. Our method for this is depicted in Figure~\ref{fig:stratspace} and embodies our intuition that most patterns of human choice in such scenarios are a near-exclusive combination of the matching/maximizing strategy when seeking, or the antimatching/minimizing strategy when hiding. Although the graph on the right of Figure~\ref{fig:stratspace} looks similar to the one on the left, their axes represent completely different things. On the left is the space of options: one axis for each. In the case of the hide-and-seek game we run experimentally, each option is a room. On the right is the space of how much maximizing ($\alpha$) and how much matching ($\beta$) are combined to make the choice vector. While the dimensionality of the histogram representation changes as the number of choice options changes in a scenario, the size of the strategy space representation is always fixed at two, because we always consider a combination of two strategies. This is an advantage for cross scenario comparison, but it is also reflects a specific and concrete prediction about the nature of the types of strategies that are used to form the individually variable choice proportions. 

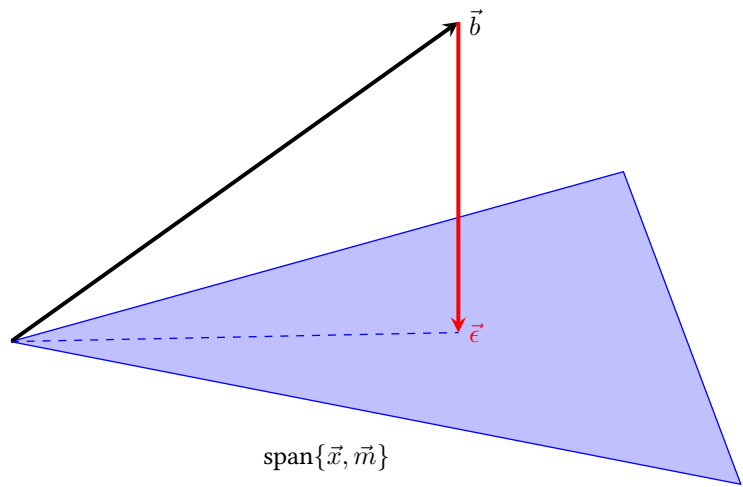
\begin{figure}[tbp]
\centering
\begin{tikzpicture}
\begin{axis}[clip=false,width=1.2*\textwidth,
                axis lines=none,
                xmin=0,
                ymin=0,
                zmin=0,
                xtick={1},
                ytick={1},
                ztick={1},
                view={40}{15}]
\addplot3[patch, patch type=triangle, color=blue, fill opacity=0.25, 
faceted color=blue, line width=0.5pt] 
coordinates{(1,0,0) (0,1,0) (0,0,0)};
\draw[line width=1.5pt, -stealth] (0,0,0)--(1/3,1/3,3/4) node[anchor = west]{$\vec{b}$};
\draw[line width=1.5pt, color=red, -stealth] (1/3,1/3,3/4)--(1/3,1/3,0) node[anchor = west]{$\vec{\epsilon}$};
\draw[line width=0.5pt, color=blue, dashed] (0,0,0)--(1/3,1/3,0) node[anchor = west]{};
\node[text width = 4cm] at (0.3,0.3, -0.3) {span$\{\vec{x}, \vec{m}\}$};
\end{axis}
\end{tikzpicture}
\caption{Testing: In higher dimensional spaces the vector representing a person's selection frequencies $\vec{b}$ (black) may not exist on the plane created by a weighted sum of the maximizing $\vec{x}$ and matching strategies $\vec{m}$ (blue region). This error vector $\epsilon$ (red) quantifies the  discrepancy, and can be used as a convenient expression of the goodness of the model.}
\label{fig:errorvec}
\end{figure}

In the two-dimensional choice case it is always possible to re-frame our two-dimensional observed choice vector as the sum of two basis strategy vectors. Thus, we will always be able to construct a participant's seek choices in a two room hide-and-seek task as a combination of maximizing and matching basis vectors, and hide choices as a combination of minimizing and antimatching basis vectors. A two-room condition thus serves as a demonstration test where it is easy to see both the raw data plotted as the frequency of choosing one room or the other. The true empirical test of the idea rests on the conditions with greater numbers of rooms (3, 5, and 7). If we can fit participant data with these two basis vectors, and not others, in the higher dimensional conditions, then we can argue that we have captured a useful decomposition for our participants' choice frequencies. Figure~\ref{fig:errorvec} shows this diagrammatically. Conceptualizing human choice proportions as vectors gives us a method of graphical representation, and a geometrical method for assessing the plausibility of hypotheses about the number of independent strategies that combine to describe peoples' choices. The closer participant choice vectors are to the plane formed by our proposed strategy combination, the better our strategy combination explains observed behavior. The observation of matching behavior and the optimality of maximizing behavior provide the principal starting points for exploring these ideas.

To recapitulate, we know that matching behavior is common in human choice experiments. We also know that maximizing choice strategies are typically optimal under the usual experimental conditions. So, we conjecture that individual choice proportions in such tasks primarily reflect for most people a combination of the two strategies. We describe this with Euclidean vector operations. We use this vector representation, and the successful history of vector based models in neural and cognitive science, to motivate our postulating a new behavioral phenomenon: probability antimatching. Probability antimatching (and minimizing) will be found when a task requires avoidance of a loss rather than acquisition of a gain. Further, the probability antimatching distribution is proposed to be the vector reflection of the probability matching vector across the uniform.

Our experiments involve a varying number of rooms for the participant to choose from. This serves two purposes. First, the greater the number of rooms for which our two-dimensional strategy account seems to fit, the more confidence we can have in our proposal of this particular low dimensional decomposition. Second, varying the number of choices probes how people might rely on these basis strategies. As the number of choice options grows so does the cognitive bookkeeping. Tracking the number of choices made for each of seven rooms requires more cognitive resources than tracking the choices for three rooms. To successfully implement an matching/antimatching policy a participant needs to track the frequencies of all choices. Intuitively, an maximizing/minimizing strategy, which only requires determining the most (or least) chosen option, could be more tolerant of imprecision, and comparatively easier to implement. We would predict an increase in the weighting of maximizing/minimizing as the number of options increases.

In the sequel we present the empirical data. We present three studies where participants played a children’s game of hide-and-seek on a computer. The game provided an intuitive way to explore both pursuit and avoidance and to assess for matching/antimatching choice patterns \cite{crawford2007fatal, chapman2014playing}.  In Experiment 1, we deploy our methodology of representing  participant strategies as a linear combination of Euclidean vectors. Using this methodology, we demonstrate participant seeking behavior is a combination of the optimal maximizing strategy and probability matching. 

We show that randomly generated surrogate participant data is not well-captured by our proposed strategies. We also show that randomly chosen strategy bases fail to capture real participant behaviour. Taken together, our proposed strategy basis is both sensitive and specific to real participant behaviour.

With these controls established, we demonstrate that hiding behavior is a combination of the optimal minimizing strategy and probability antimatching. We show that when the problem complexity grows (from a 2 room to 5 room scenario in these experiments) the participant strategy mixes differ with hide/seek context. In Experiment 2, we replicate Experiment 1 for online data collection. In Experiment 3, we present new probability distributions, some of which have reflections that fall outside the probability simplex, a phenomeon we will explain in detail in the introduction to Experiment 3.Experiment 3 again will demonstrate our expected hide/seek strategy mixes. We also show how the two projection methods for recovering from illegal reflections to the simplex give good approximations to participant behavior.

\section{Methods}
\subsection{Behavioral}
We developed a computerized version of the children's game ``hide-and-seek''. Each participant played a set of hide-and-seek games against simulated \emph{predictable} opponents. They were predictable in that they hid and sought with a fixed probability distribution. This was known to all participants. We provided trial by trial counters to provide additional evidence to the participants that the information was accurate. 

\begin{figure}[tbp]
\centering
  \includegraphics[width=0.7\textwidth]{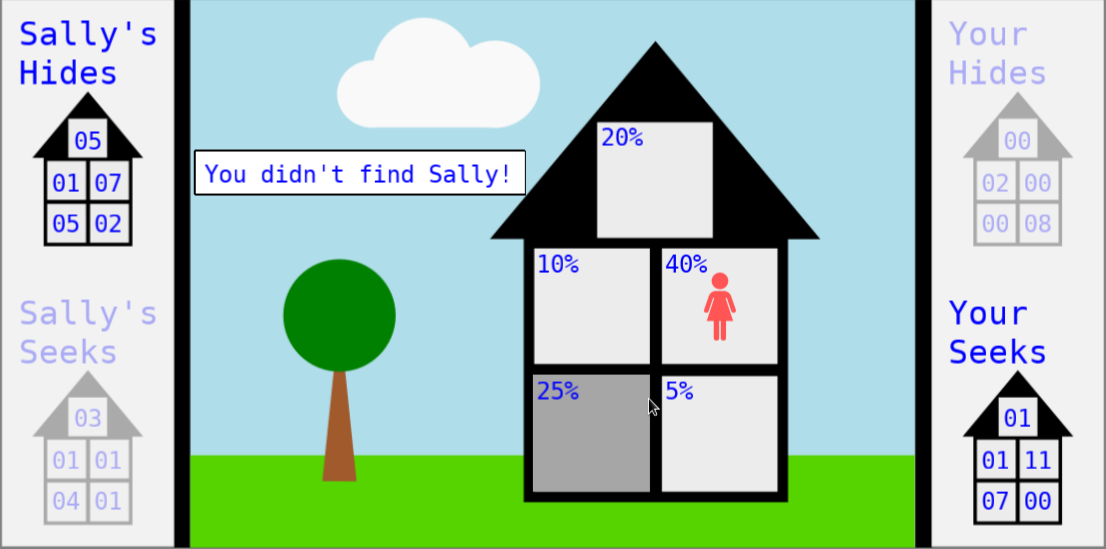}
  \caption{Participant view of screen: This is the five-room condition. After selecting the bottom left room in a seek trial the selected room turned grey. The child is then revealed, a notification is presented, and the counters are updated.}
  \label{fig:pythonhouse}
\end{figure}

Our user interface (Figure~\ref{fig:pythonhouse}) consisted of a cartoon scene of a house with either two, three, five, or seven rooms. The house was of a `doll house style' where each room was visible to the participant. We presented a percentage within each room. Each percentage represented the probability that the opponent would hide/seek in that room. Room proportions always summed to 100\%. Participants played against a fixed set of distributions, but the mapping between rooms and proportions for a given distribution were randomized for each participant. Presenting the actual choice counts to participants allowed us to attribute performance to their strategy construction and not ambiguity or imprecision in their estimates of opponent choices or memory for their own past choices. We were not interested in the learning of the probabilities, but in how the probabilities were used to guide their choices. 

\subsection{Task Procedure}
Each experiment began with a short practice round consisting of three seeking trials followed by three hiding trials. As in standard hide-and-seek games, each participant was instructed to find each opponent while seeking, and avoid each opponent while hiding.

A dialogue box notified the participant that the child had hidden. The screen displayed ``Look for [child]!'' (where [child] is the name of the current opponent). Using the mouse the participant selected a room to search. Afterwards, either a notification of success (``You found [child]!'') or a notification of failure  (``You didn’t find [child]!'') was displayed. As the hider, the dialogue box read ``Hide from [child]!'' The participant selected a room to hide using the mouse. Similarly, either ``[child] didn’t find you!'' was displayed on screen to indicate success, or ``[child] found you!'' to indicate failure. The child was always revealed after any successful or failed attempt for both trial types. 

A single hide or seek choice of the participant was treated as a `trial.' The child’s hiding and seeking locations were drawn from the distribution displayed over the rooms and were independent between trials. A series of 10 trials constituted a `round.' The experiments alternated between seeking and hiding rounds, always beginning with seeking. A set of 10 seeking rounds and 10 hiding rounds against a particular child constituted a `game.' Participants played a total of three games in Experiments 1 and 2 and four games in Experiment 3. To prevent frustration and boredom from interfering with performance the game where the opponent always hid or sought in the same one room was played last.

\newcommand{\dummyfigure}{\tikz \fill [NavyBlue] (0,0) rectangle node [black] {Figure} (2,2);}
\newcolumntype{M}[1]{>{\centering\arraybackslash}m{#1}}
    \begin{table}
        \centering
        \begin{tabular}{cM{20mm}M{20mm}M{20mm}}
           \toprule
            Opponent name (order) & 2 Room & 5 Room \\
            \midrule
            \includegraphics{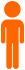}
            Toby (Practice) &
            \resizebox{20mm}{20mm}{%
            \begin{tikzpicture}
              \begin{axis}[clip=false,ybar interval, ymax=1,ymin=0, minor y tick num = 3]
                \addplot coordinates { (1, 0.6) (2, 0.4) (3, 0) };
              \end{axis}
            \end{tikzpicture}
            } &
            \resizebox{20mm}{20mm}{%
            \begin{tikzpicture}
              \begin{axis}[clip=false,ybar interval, ymax=1,ymin=0, minor y tick num = 3]
                \addplot coordinates { (1, 0.3) (2, 0.3) (3, 0.2) (4, 0.1) (5, 0.1) (6, 0) };
              \end{axis}
            \end{tikzpicture}
            }  \\
            \includegraphics{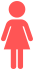}
            Sally (1st or 2nd)
            & \resizebox{20mm}{20mm}{%
            \begin{tikzpicture}
              \begin{axis}[clip=false,ybar interval, ymax=1,ymin=0, minor y tick num = 3]
                \addplot coordinates { (1, 0.7) (2, 0.3) (3, 0) };
              \end{axis}
            \end{tikzpicture}
            } &
            \resizebox{20mm}{20mm}{%
            \begin{tikzpicture}
              \begin{axis}[clip=false,ybar interval, ymax=1,ymin=0, minor y tick num = 3]
                \addplot coordinates { (1, 0.4) (2, 0.25) (3, 0.2) (4, 0.1) (5, 0.05) (6, 0) };
              \end{axis}
            \end{tikzpicture}
            } \\
            \includegraphics{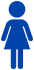}
            Kala (1st or 2nd)
            & \resizebox{20mm}{20mm}{%
            \begin{tikzpicture}
              \begin{axis}[clip=false,ybar interval, ymax=1,ymin=0, minor y tick num = 3]
                \addplot coordinates { (1, 0.5) (2, 0.5) (3, 0) };
              \end{axis}
            \end{tikzpicture}
            } &
            \resizebox{20mm}{20mm}{%
            \begin{tikzpicture}
              \begin{axis}[clip=false,ybar interval, ymax=1,ymin=0, minor y tick num = 3]
                \addplot coordinates { (1, 0.2) (2, 0.2) (3, 0.2) (4, 0.2) (5, 0.2) (6, 0) };
              \end{axis}
            \end{tikzpicture}
            } \\
            \includegraphics{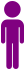}
            Bo (3rd)
            & \resizebox{20mm}{20mm}{%
            \begin{tikzpicture}
              \begin{axis}[clip=false,ybar interval, ymax=1,ymin=0, minor y tick num = 3]
                \addplot coordinates { (1, 1) (2, 0) (3, 0) };
              \end{axis}
            \end{tikzpicture}
            } &
            \resizebox{20mm}{20mm}{%
            \begin{tikzpicture}
              \begin{axis}[clip=false,ybar interval, ymax=1,ymin=0, minor y tick num = 3]
                \addplot coordinates { (1, 1) (2, 0) (3, 0) (4, 0) (5, 0) (6, 0) };
              \end{axis}
            \end{tikzpicture}
            } \\
            \bottomrule
        \end{tabular}
        \caption{Opponents faced by participants in each condition of Experiments 1 and 2. Each distribution is presented here in decreasing room
probabilities, though probabilities were randomly assigned to rooms during the experiment.}
        \label{tbl:pythonchars}
    \end{table}

The experiment was programmed in Python for Experiment 1 using the PsychoPy module \cite{peirce2019psychopy2} and completed in-lab. Due to the COVID-19 pandemic, Experiments 2 and 3 were conducted online. We reproduced the task using JsPsych \cite{de2015jspsych}. This JavaScript implementation allowed for online data collection. Other than a slight visual change in the house due to the different method of rendering the image, the only other procedural change was to have participants advance from one trial to the next by clicking a button on the screen rather than using the space-bar of the keyboard as was used in Experiment 1. 

At the end of Experiments 1 and 2, participants answered demographic questions of gender, age, term of study, and academic program, along with their perceived competence in logical reasoning relative to other students in that program.

Despite being generally similar in all the procedural aspects of the task, Experiment 3 had a more complicated structure of rooms and distributions. The distribution each participant faced in the first game was randomly selected from one of three sets: 3-, 5-, or 7-room distributions. The second game distribution was randomly selected from the remaining two sets that did not contain the distribution selected for the first game. The third distribution was randomly selected from the remaining set. The fourth game was randomly selected from a set that only contained distributions with probability mass all in one room.

The distributions and the selection procedure can be efficiently outlined with a vector notation. The symbol $\in_R$ denotes `randomly selected from'. A backslash in set notation denotes exclusion. For example, $\{A,B,C,D\} \backslash \{A,C\} = \{B,D\}.$

Let

  \begin{flalign*}
    D^3 &= \left\{
    \begin{bmatrix}
           50 \\
           25 \\
           25
    \end{bmatrix},
    \begin{bmatrix}
           42 \\
           42 \\
           16
    \end{bmatrix},
     \begin{bmatrix}
           80 \\
           10 \\
           10
    \end{bmatrix}
    \right\},&&
  \end{flalign*}

  \begin{flalign*}
    D^5 &= \left\{
    \begin{bmatrix}
           35 \\
           30 \\
           15 \\
           15 \\
           5
    \end{bmatrix},
    \begin{bmatrix}
           35 \\
           25 \\
           25 \\
           10 \\
           5
    \end{bmatrix},
     \begin{bmatrix}
           45 \\
           35 \\
           10 \\
           5 \\
           5
    \end{bmatrix}
    \right\},&&
  \end{flalign*}

  \begin{flalign*}
    D^7 &= \left\{
    \begin{bmatrix}
           25 \\
           25 \\
           20 \\
           14 \\
           10 \\
           4 \\
           2
    \end{bmatrix},
    \begin{bmatrix}
           26 \\
           24 \\
           18 \\
           15 \\
           9 \\
           4 \\
           4
    \end{bmatrix},
     \begin{bmatrix}
           50 \\
           18 \\
           12 \\
           8 \\
           5 \\
           5 \\
           2
    \end{bmatrix}
    \right\}, \text{and} &&
  \end{flalign*}

  \begin{flalign*}
    D' &= \left\{
    \begin{bmatrix}
           100 \\
           0 \\
           0 
    \end{bmatrix},
    \begin{bmatrix}
           100 \\
           0 \\
           0 \\
           0 \\
           0 
    \end{bmatrix},
     \begin{bmatrix}
           100 \\
           0 \\
           0 \\
           0 \\
           0 \\
           0 \\
           0
    \end{bmatrix}
    \right\}.&&
  \end{flalign*}

\begin{itemize}
  
\item Each participant plays 4 games: $G_1, G_2, G_3, G_4$
  
\item Each game $G_i$ has one opponent that follows a particular hide-and-seek distribution, $\vec{g_i}$

\item For any game $G_i$ with distribution $\vec{g_i}$, each element of $\vec{g_i}$ is randomly assigned to each room of the house presented on the screen

\item $\vec{g_1} \in_R D^n$, where $n \in_R \{3,5,7\}$

\item $\vec{g_2} \in_R D^m$, where $m \in_R \{3,5,7\} \backslash \{n\}$

\item $\vec{g_3} \in_R D^k$, where $k \in_R \{3,5,7\} \backslash \{n, m\}$

\item $\vec{g_4} \in_R D'$

\end{itemize}

\subsection{Participants}

We recruited a total of 281 University of Waterloo students to participate in exchange for a course credit. All participants gave informed consent and the study was cleared by a University of Waterloo Research Ethics Board (REB 41316). Experiments 1, 2, and 3 took approximately 25, 21, and 32 minutes to complete, respectively. For Experiment 1 there were 50 participants (38 female). For Experiments 2 there were 54 participants (44 female) and for Experiment 3 there were 177 participants (134 female). No participants declared any other genders than male or female. 

\subsection{Data Analysis}
This section elaborates on the analytical procedures used to quantify the ideas outlined in the introduction: that people use frequency information as the primitive for constructing  a small set of canonical reference strategies that are mixed to yield their selection policy. The frequency information that participants experience is of dimension $N$ where $N$ is the number of rooms in which hiding or seeking is possible. While any point on the $N$-dimensional probability simplex could, in principle, yield participant choice frequencies, we proposed that two basis strategies are mixed to yield participant choices. The basis strategies are either maximizing/matching when seeking or minimizing/antimatching when hiding. The span of these two input histograms gives us a 2-dimensional linear sub-space. For any participant choice behavior, we can define a 'best fit' representation within this subspace. Any imprecision between best fit within the subspace and a participant's actual choice behavior can be summarized by the error vector we get from subtracting our best fit from the participant's choice histogram. The magnitude of that vector is a proxy measure of goodness of fit (Figure~\ref{fig:errorvec}). 

All participants knew the reference distribution. It was shown to them, and it was confirmed by the simulated child's choices, which was tracked with counters. This distribution provides the theorized matching distribution. The maximizing distribution was constructed by putting a mass of one on the room with the highest probability of being chosen by the child. The minimizing distribution was constructed by putting a mass of one on the room least often chosen. The antimatching strategy was calculated from the matching distribution by reflection across the uniform, except for some slight subtleties to be presented in the results section of Experiment 3.

A first analysis looked for the expected pattern of probability matching by inspecting visually the histogram of participant choices. For seeking games, we plotted all participant seek frequencies overlaid with the opponent's hide frequencies. Similarly, for hiding games, we plotted all participant hide frequencies overlaid with the opponent's seek frequencies. We calculated the best linear combination of our two theorized basis strategies using the Stark-Parker algorithm \cite{stark1995bounded} for bounded-variable least squares via the ‘bvls’ package in R \cite{bvls}. We used bounded least squares to ensure our model coefficients were between 0 and 1 (inclusive) because no strategy can be used less than 0\% of the time or more than 100\% of the time. Variables and abbreviations are summarized in Table~\ref{table_vars}.

\begin{table}[tbp] 
    \centering
    \begin{tabular}{lp{9cm}}
        \toprule
        \textbf{Variable} & \textbf{Definition}  \\
        \midrule
        $\vec{b}$ & Recorded participant behavior \\
        $\vec{x_{s|h}}$ & Maximizing or minimizing strategy \\
        $\vec{m_{s|h}}$ &  Matching or antimatching strategy \\
        $\alpha_{s|h}$ & Amount of $\vec{x}$ needed to best describe $\vec{b}$ \\
        $\beta_{s|h}$ & Amount of $\vec{m}$ needed to best describe $\vec{b}$ \\
        $\vec{\epsilon}$ & Error in representing $\vec{b}$ with $\vec{x}$ and $\vec{m}$ \\
        \bottomrule
    \end{tabular}
     \vskip.1in\par
     {\textit{Note}: Vectors and parameters with the subscript \textit{r} are randomly generated for sensitivity analyses. $\vec{u}$ denotes the uniform distribution.}
    \caption{Definition of variables. Subscripts indicate ``(s)eeking'' or ``(h)iding''. }
    \label{table_vars}
\end{table}

\subsubsection{Accounting for infinite optimal strategies}

In Experiments 1 and 2, the optimal strategies were unique. For Experiment 3 it was necessary to account for the fact that some distributions had an infinite number of optimal strategies. This occurs for seeking when the maximum probability room is not unique, and for hiding where the minimum probability room is not unique. Interestingly, if there is no unique optimal strategy, then there are theoretically an infinite number of optimal strategies if participants could make an infinite number of choices, but practically since our participants only completed 100 trials there are finitely many optimal choice frequencies. Our analysis accommodated the infinite theoretical case so as to make this analysis generalizable for future work.

For any distribution that has a non-unique maximum or minimum room, the maximizing and minimizing strategies are generalized from the vectors $x_s$ and $x_h$, respectively, to the sets of vectors $X_s$ and $X_h$ , respectively. The number of vectors in $X_s$ is determined by the number of the rooms that share the maximum probability value. Analogously, the number of vectors in $X_h$ is determined by the number of rooms that share the minimum probability value. Both $X_s$ and $X_h$ contain only standard unit vectors, denoted $e$ , where  the \emph{i}’th element is 1 and all  other elements are 0. The set $X_s$ ($X_h$) contains vectors $e$, where \emph{i} is selected from the room numbers that have the maximum (minimum) probability value. For example, if $\vec{m_s}$ = (0.42, 0.42, 0.16), then $X_s$ = \{(1, 0, 0), (0, 1, 0)\}. If $\vec{m_h}$ = (0.26, 0.24, 0.18, 0.15, 0.09, 0.04, 0.04), then $X_h$ = \{(0, 0, 0, 0, 0, 1, 0), (0, 0, 0, 0, 0, 0, 1)\}.

\section{Results Experiment 1 (in-lab) Two Room and Five Room (Sally)}

We begin with the experiment that comes closest to the traditional two-alternative forced choice procedure that has been used to demosntrate probability matching. This presents our distinct approach to data visualization and analysis in a familiar setting and serves as a basic proof of concept, though it is only in the conditions with higher numbers of rooms that we can begin to evaluate the claim that a linear combination of two particular strategy vectors does an excellent job of accounting for participant choices. A plot of the frequency with which each of our participants searched against ``Sally'' in  the two room condition confirms the predominance of probability matching in our cohort, though a few participants did use the optimal, maximizing, strategy, and others showed some mix of strategies. (Figure \ref{fig:s2}) also reveals the context sensitivity. Participants chose differently when hiding. Even though participants have identical probability information in the hide and seek conditions matching is replaced with antimatching, and maximizing is replaced with minimizing.

\begin{knitrout}
\definecolor{shadecolor}{rgb}{0.969, 0.969, 0.969}\color{fgcolor}\begin{figure}
\includegraphics[width=\maxwidth]{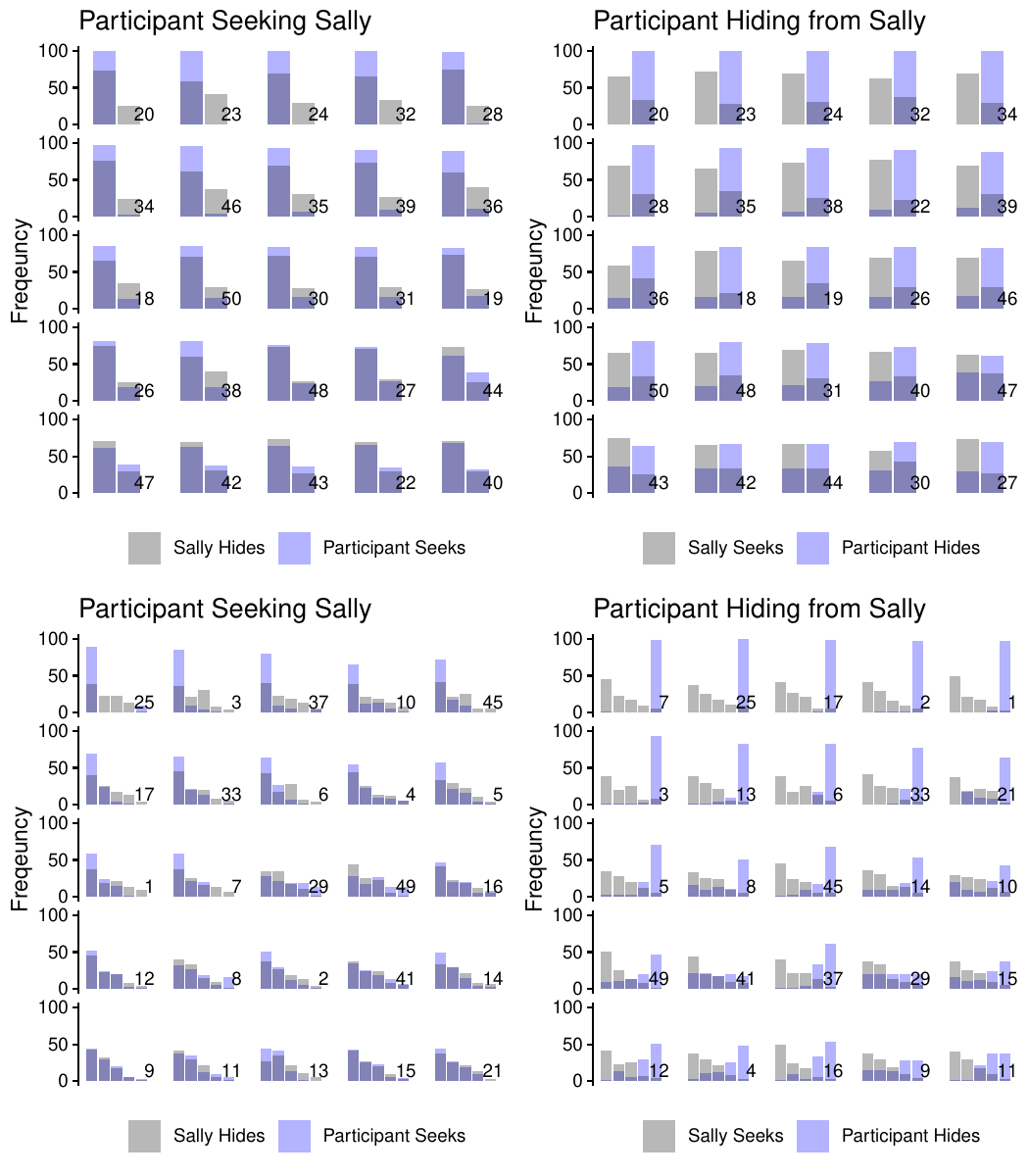} \caption[Frequency histograms for participants seeking/hiding with the Sally opponent in the two room (top row) and five room (bottom row) scenarios]{Frequency histograms for participants seeking/hiding with the Sally opponent in the two room (top row) and five room (bottom row) scenarios: The left column shows histograms for seek choices, and the right column for hide choices. The participant numbers are arbitrary, but consistent across panels. The gray bars show programmed opponent room frequencies. Darker areas are areas of overlap. Participants are plotted in order of strategy use, with mostly maximizing/minimizing in the top left of each panel, and mostly matching/antimatching in the bottom right of each panel. When seeking, matching produces entirely dark bars.}\label{fig:s2}
\end{figure}

\end{knitrout}

\subsection{Comparing Hiding and Seeking}

While Figure~\ref{fig:s2} gives the conventional data summary we can also re-plot participant data in the low dimensional strategy space. This alternative visualization collapses much of the data to a single representation. The alternative visualization begins by setting the strategy space to be that defined by the maximizing/minimizing and matching/antimatching strategies (Figure~\ref{fig:s2Strat}). This strategy space depicts $\alpha$ on the y axis and $\beta$ on the X axis. Any participant strategy that uses exclusively a maximizing/minimizing strategy will appear as a point placed in the top left of the plot (0,1). Any participant that uses exclusively a matching/antimatching strategy will appear in the bottom right of the plot (1,0). Any strategy that is made exclusively of some mix of maximizing/minimizing  and matching/antimatching will appear on the diagonal line connecting (0,1) and (1,0). Whether that two-dimensional representation is sufficient is an empirical question.

\begin{knitrout}
\definecolor{shadecolor}{rgb}{0.969, 0.969, 0.969}\color{fgcolor}\begin{figure}
\includegraphics[width=\maxwidth]{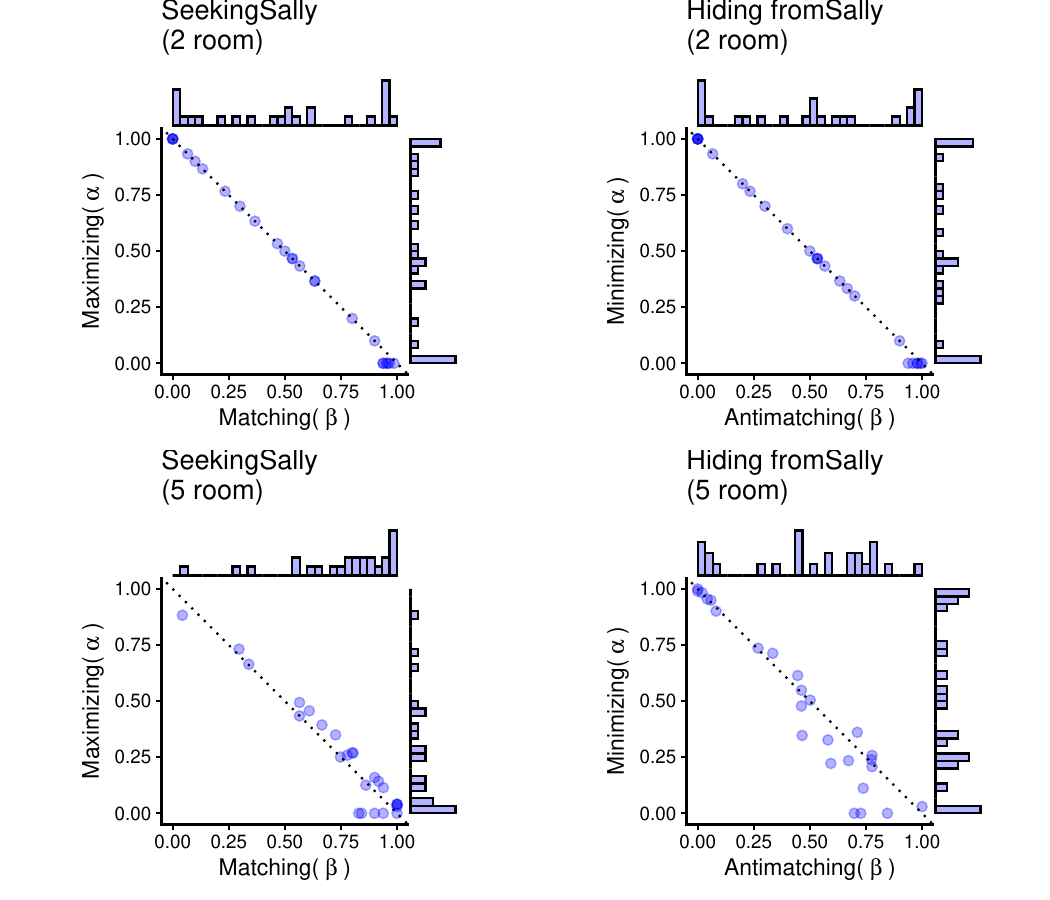} \caption[Strategy space plot]{Strategy space plot: We fix two basis distributions: maximizing/minimizing and matching/antimatching. We determine for each participant the optimal combination of these two vectors that gets as close as possible to their choice frequencies (in the two dimensional case we can almost do this perfectly). We label the maximizing/minimizing weight as $\alpha$ and the matching/antimatching as $\beta$. Each participant is a point in this space. For each panel, the closer they are to the upper left the more they maximizing/minimizing. The closer to the lower right the more they matching/antimatching. The proximity to the line of slope -1 gives a measure of how well we can do approximating choice performance with just these two basic strategies (all participants lie on or essentially on the line $\alpha + \beta = 1$ implying there is negligible error in this fit.}\label{fig:s2Strat}
\end{figure}

\end{knitrout}

Figure~\ref{fig:s2-5SallyChangeStratVec} shows the two room and five room Sally data. Nothing forces these change vectors to lie on the diagonal line from (0,1) to (1,0), in the five-room case, but we observe this to be true since most seeking strategies are an exclusive mix of maximizing and matching, and most hiding strategies are an exclusive mix of minimizing and antimatching. Moreover, we see that the average strategy mix is closer to 0,1 when hiding than when seeking, but only in the five-room condition. This indicates that in the five-room condition, participants generally use a greater mix of the optimal strategy when hiding than when seeking.

\begin{knitrout}
\definecolor{shadecolor}{rgb}{0.969, 0.969, 0.969}\color{fgcolor}\begin{figure}
\includegraphics[width=\maxwidth]{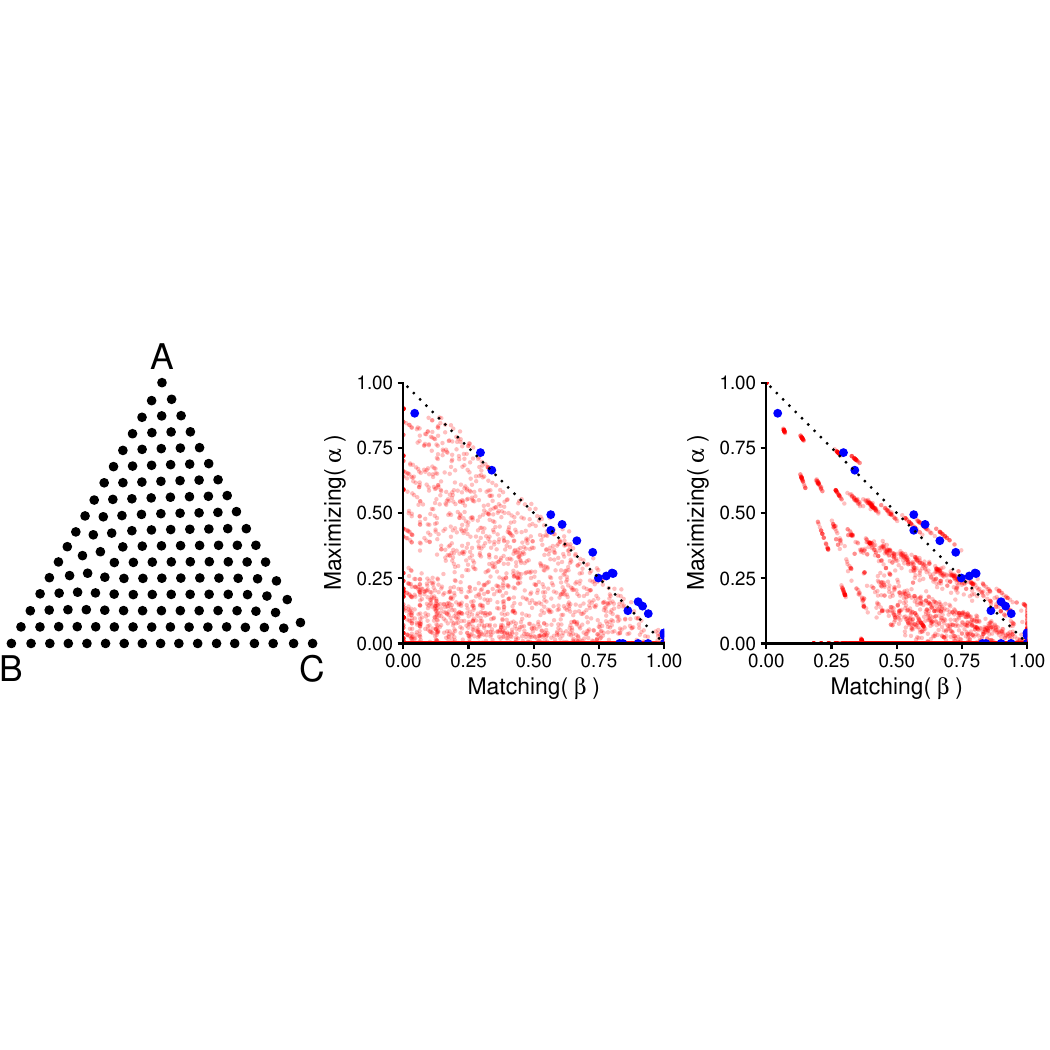} \caption[Surrogate Data Analysis]{Surrogate Data Analysis. We used the five room condition with the "Sally" distribution to test the account that the strategy space plots are an artifact. The surrogate data sampling procedure resulted in data uniformly distributed across the probability simplex. A three dimensional surrogate data set is shown here as an illustration. A, B, C represent possible hiding and seeking locations. An $N$ choice options yields an $N ~ - ~ 1$ simplex. The middle panel shows the strategy space plot for each surrogate distribution used as the matching (and corresponding maximizing) as the strategy space for each participant. The original empirical fits are shown in blue. The right panel shows the fits one gets when using our original matching and maximizing vectors as the strategy basis and instead treats each surrogate as potential participant response (red). Blue dots (same as middle panel) are the empirical data fits and are included for comparison. In summary, the empirical fits in strategy space, and their propensity to align around the alpha + beta = 1 line is not due to an arbitrary low dimensional representation or a particularly fortunate choice of strategy basis set pair. Our theoretically motivated strategy basis is both sensitive and specific to real participant data.}\label{fig:surrogatePlotsE1}
\end{figure}

\end{knitrout}

The five-room data provide the first real test of the idea that a pair of basis strategies from maximizing/minimizing and matching/antimatching can provide a good fit for human choice behavior, because in general, two independent vectors can not cover a five dimensional space. The pattern is similar to the two-room data: Both hiding and seeking strategies are well represented by a combination of maximizing/minimizing and matching/antimatching (Figure~\ref{fig:s2} and Figure~\ref{fig:s2Strat})

To show that this is not accidental or an artifact we consider two alternatives: first, that it just so happens that the strategies people employ in tasks like ours are inherently two dimensional, and thus any two independent vectors would do just as well as the pair we chose, or second, that we got lucky and picked two ``special'' vectors that would have fit well any  possible participant choice frequency. For both alternatives we tested with a method of surrogate data. We generated 250 points on the probability simplex of a five dimensional space. These points were uniformly distributed across the simplex. These points were computed with the \texttt{get\_reference\_directions} function of the \texttt{pymoo} Python library \cite{pymoo}.  To illustrate the spacing of the data Figure~\ref{fig:surrogatePlotsE1} (left) shows a surrogate set for a three dimensional problem (three dimensions is used so that we can plot it). For testing the first alternative we took each of the 250 five-dimensional vectors and treated it like a matching distribution. We generated the corresponding surrogate maximizing distribution and calculated the coefficients for each using the same code we used for the empirical data. We also recorded the deviance, that is the magnitude of the vector from the fit point in strategy space to the true participant's choice frequencies. We used the data from the five room ``Sally'' condition. Figure~\ref{fig:surrogatePlotsE1} (middle) shows the surrogate fits in red and the empirical points in blue. Although the failure of surrogate data to do a good job of fitting the empirical data seems visually clear we compared them statistically. We took the median deviance for each surrogate and the deviance for the true maximizing and matching distributions and compared them with a non-parametric test, the Wilcoxon Rank Sum. The deviance of the surrogates generated a statistically worse fit than did the true frequencies (Wilcoxon Rank Sum W = 0 p value = 0. The extremity of this p-value is because there was no case where a median deviance for the surrogate was less than a deviance for any of the true pairs. Thus, it is not the case that a random vector from the probability simplex (analogous to our matching distribution) with a corresponding paired maximum surrogate  would be expected to do as well as our theoretically motivated choice.

To consider the alternative that for some coincidental reason we just happened to hit on a powerful pair of basis strategies we kept our original strategies as the references and fit each of the surrogates as if it had been some participant's choice frequency.  This is also seen to be false (Figure~\ref{fig:surrogatePlotsE1}) right). These analyses contradict the idea that participants' responses are non-specifically low dimensional or that in a probability task like ours the stimulus distribution would provide a good account of any mathematically permissible choice frequency a participant could potentially exhibit.

For comparing the strategy choices as a function of whether participants were hiding or seeking, we can extend our vector based approach to visualizing this as a change vector. For each participant we can compute this vector. If there is no systematic change between conditions, then the distribution of these change vectors should be heterogeneous in direction and average out to be of  small magnitude. Also, if, as we propose, participants' choice frequencies are a linear combination of matching/maximizing or antimatching/minimizing then we should expect to see these change vectors align along the same axis of upper left to lower right. We see in general that strategy space change does tend to move along this axis with a bias to shift toward the maximizing pole when participants are choosing their hiding places (Figure~\ref{fig:s2-5SallyChangeStratVec}).

\begin{knitrout}
\definecolor{shadecolor}{rgb}{0.969, 0.969, 0.969}\color{fgcolor}\begin{figure}
\subfloat[Two Room\label{fig:s2-5SallyChangeStratVec-1}]{\includegraphics[width=.45\linewidth]{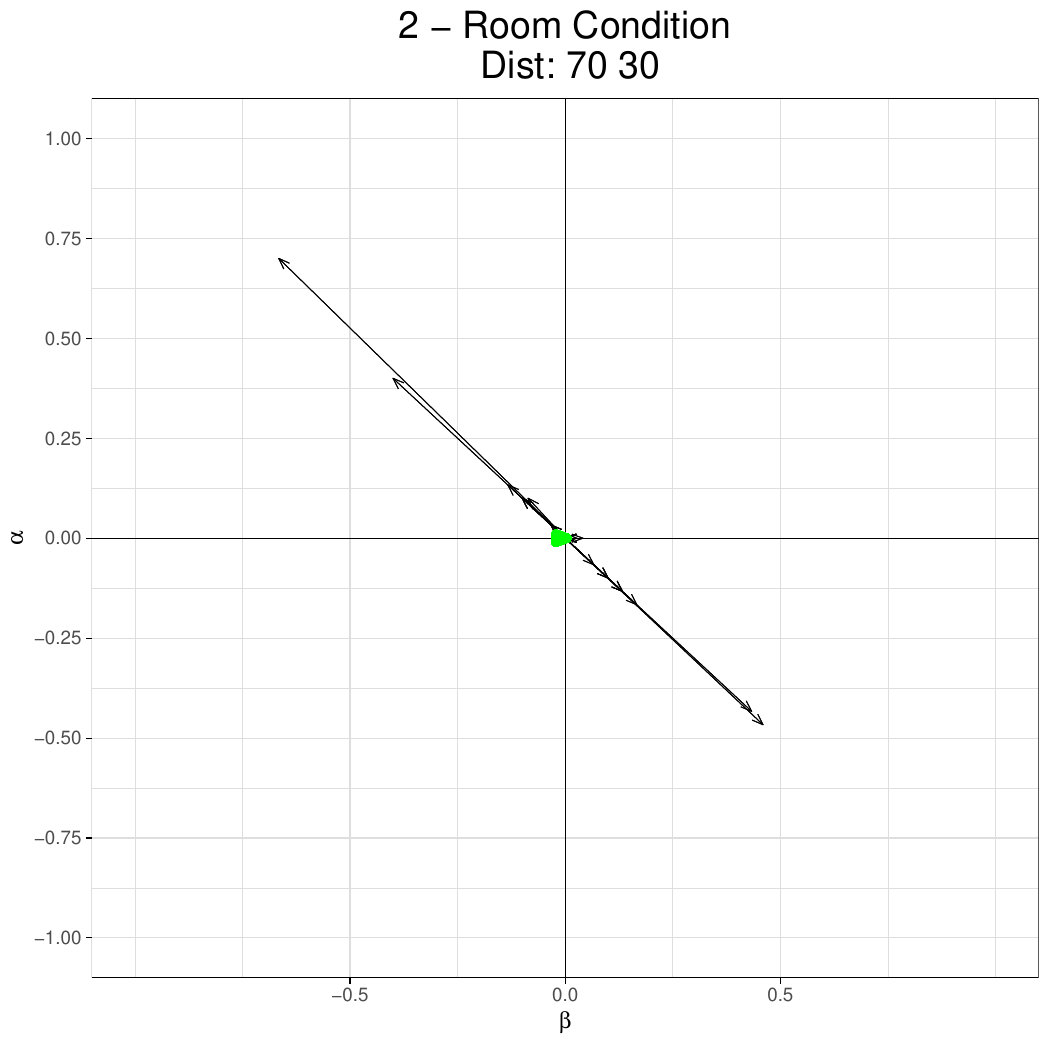} }
\subfloat[Five Room\label{fig:s2-5SallyChangeStratVec-2}]{\includegraphics[width=.45\linewidth]{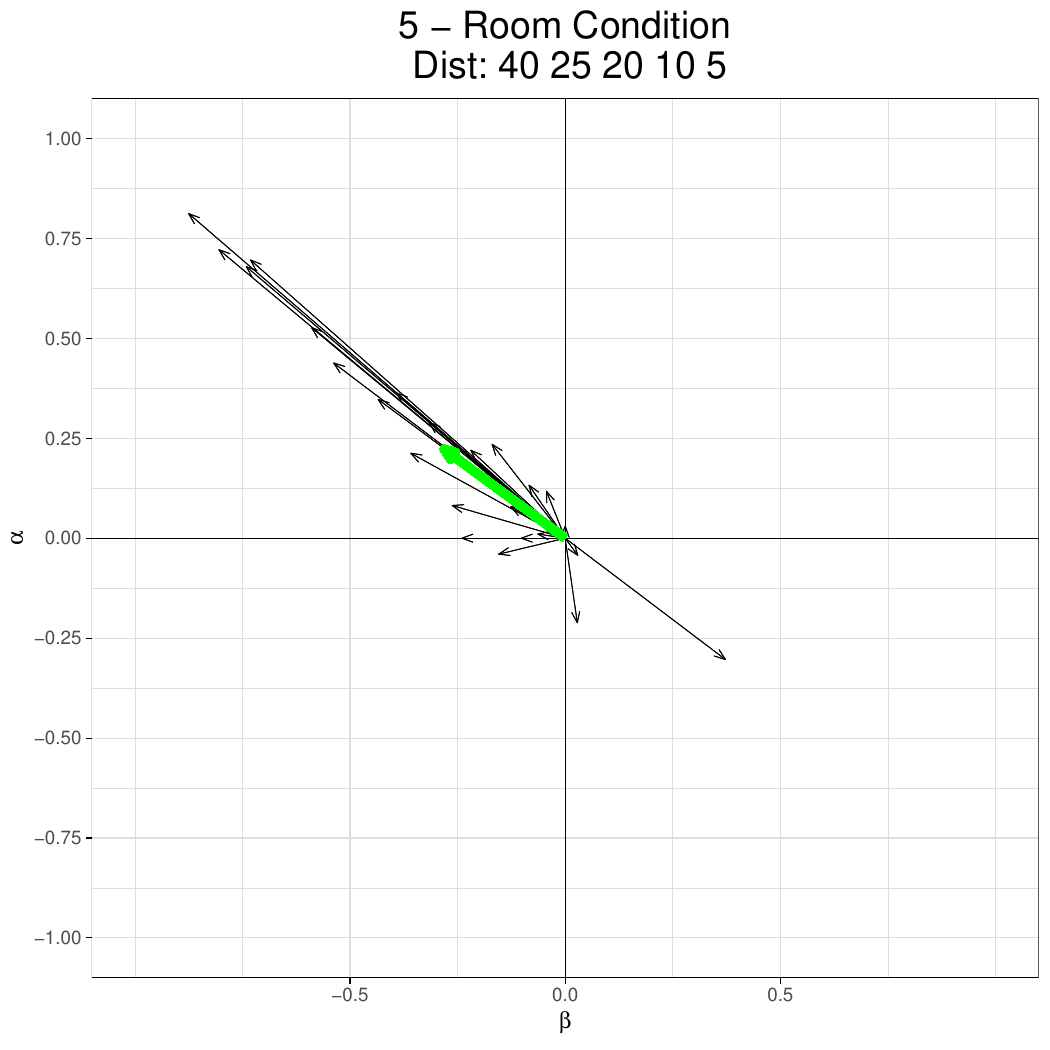} }\caption[Plotting changes in strategy vectors for the two and five room conditions (Sally distribution)]{Plotting changes in strategy vectors for the two and five room conditions (Sally distribution): The left plot shows the two-room condition and the right plot shows the five-room condition. For each plot the choice data for each participant was fit to best explain their choice data using a basis vector of the maximizing/minimizing and matching/antimatching strategies. These two vectors are subtracted (hiding - seeking) to yield a change vector that originates at the observed seeking strategy, and points to the observed hiding strategy. They are then shifted to a common starting point to highlight the magnitude and direction of change. Change vectors of individual participants are shown as thin black lines. The average of all change vectors is shown as a thick green line.}\label{fig:s2-5SallyChangeStratVec}
\end{figure}

\end{knitrout}

The principle finding of Experiment 1 is that both seeking and hiding are well represented by a combination of only two basis vectors (maximizing/minimizing and matching/antimatching) even for a condition (five rooms) that has considerably more degrees of freedom. We also find a greater use of minimizing distribution than maximizing distribution (the exploitation strategies), but only in the 5-room condition.

\subsection{Additional Distributions}

In the first experiment participants not only played against Sally, but against two other players who either chose uniformly or chose one room exclusively (Table~\ref{tbl:pythonchars}). These distributions are special. In the uniform case (Kala) it does not matter what strategy the participant employs: they will always theoretically win/lose the same number of rounds, and our plotting of strategy change vectors should be vacuous in this case. For the case (Bo) where one room is picked exclusively we should see a consistent change from only picking the 100\% room (when they are seeking) to never picking that room (when they are hiding). Our strategy change plots confirm that participants are behaving as expected in these two degenerate cases (Figures \ref{fig:e1HistsAltDistKala}  and \ref{fig:e1HistsAltDistBo}). These extra distributions probe the ambiguous edge-cases of our model, and serve as checks on the attentiveness of our participants, and as a check on our computer code. We will not be presenting further analyses of these distributions in the subsequent cases, although both distributions were included in all experiments.

\section{Results Experiment 2 - Online Replication}

\subsection{Sally: Two Room and Five Room Conditions}

Our first hide-and-seek experiment was done in person. Due to the pandemic we needed to transition to on-line administration. This experiment evaluated if our results replicated in an on-line setting. 

\begin{knitrout}
\definecolor{shadecolor}{rgb}{0.969, 0.969, 0.969}\color{fgcolor}\begin{figure}
\includegraphics[width=\maxwidth]{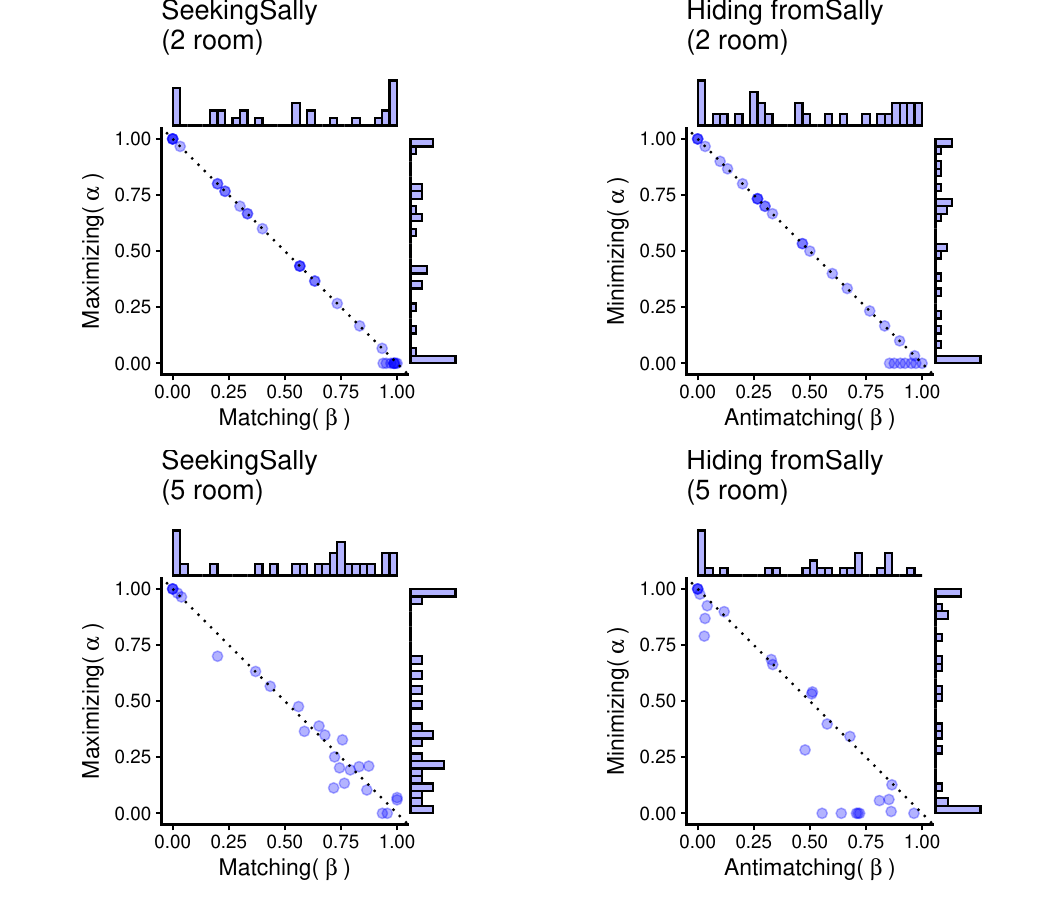} \caption[Strategy space of participant seeks against Sally]{Strategy space of participant seeks against Sally. Bars along margin count participants using particular value of $\alpha$ or $\beta$. Perfect maximizers use $\alpha$ = 1 and $\beta$ = 0 (top left). Perfect matchers use $\alpha$ = 0 and $\beta$ = 1 (bottom right). Participant strategies existing on or near line $\alpha$ + $\beta$ = 1 implies near perfect model fit.}\label{fig:osally2seekStrat}
\end{figure}

\end{knitrout}

For comparison to Experiment 1 (Figure \ref{fig:s2Strat}) the strategy space plots of $\alpha$ and $\beta$ are shown in Figure \ref{fig:osally2seekStrat}. This is largely a replication of Experiment 1 with many participants showing a preference for matching behavior when seeking. However, there is a mix of the two strategies, as before. The two strategies together provide a good account of the participants' choices (as they should for a two room condition; all participants lie on or essentially on the line $\alpha + \beta = 1$).

For the five room condition (shown in the bottom of Figure \ref{fig:osally2seekStrat}) we see broadly similar results to Experiment 1 and to the predilection for participants' strategy mix to stay close to and walk along the theoretically proposed maximizing/minimizing - matching/antimatching axis between (0,1) and (1,0) when being tested in hiding or seeking conditions. The change in the location of the ``blue dots'' between conditions and away from the theoretically proposed axis suggests that there is error in expressing our theoretically proposed strategy mix, or alternatively, other strategies are being used, but only contribute a small and inconsistent influence to participant performance. We do note though that we find more noise for Experiment 2 than Experiment 1. Since Experiment 2 was online it may be that these participants were less attentive or motivated, but ultimately the source of differences between in lab and online versions remain conjectural.

\begin{knitrout}
\definecolor{shadecolor}{rgb}{0.969, 0.969, 0.969}\color{fgcolor}\begin{figure}
\subfloat[Two Room\label{fig:s2-5SallyChangeStratVecE2-1}]{\includegraphics[width=.45\linewidth]{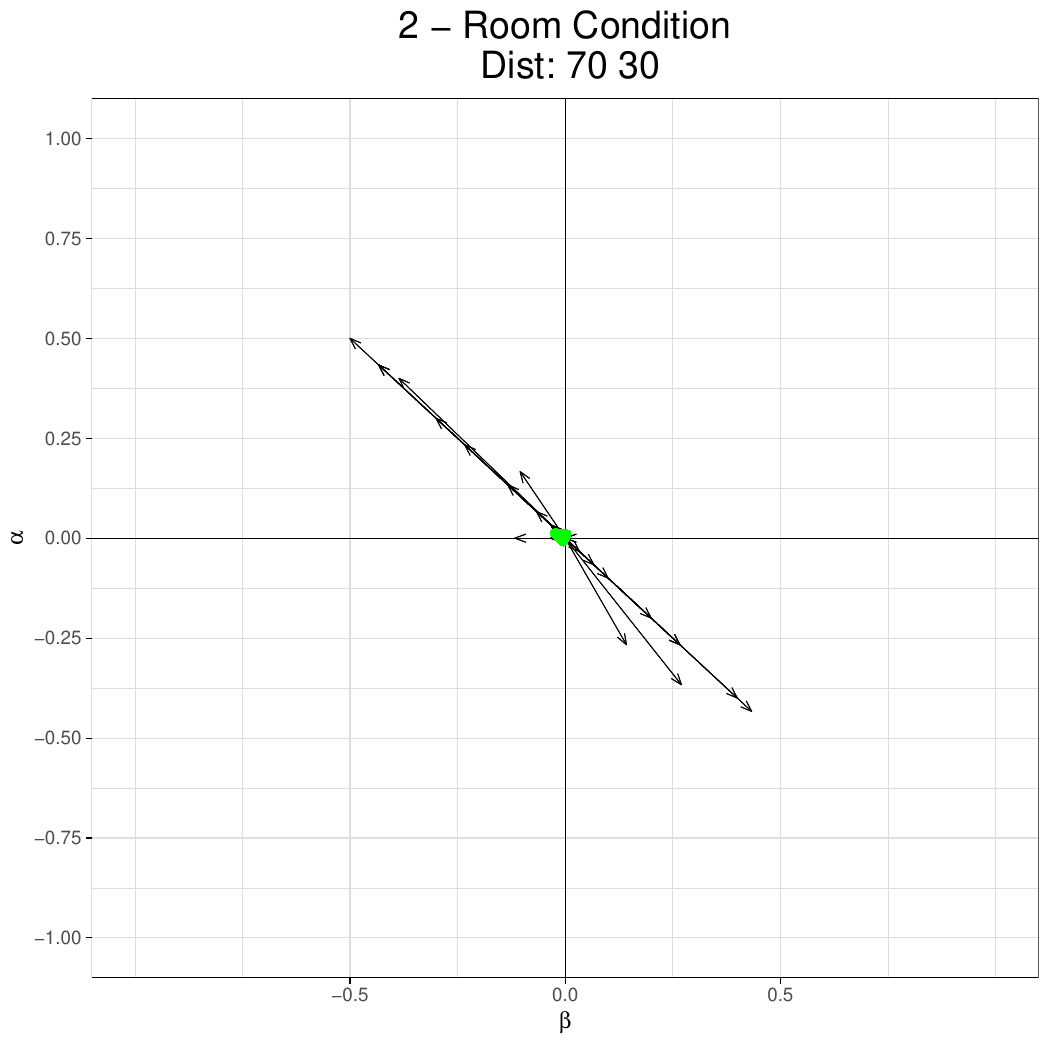} }
\subfloat[Five Room\label{fig:s2-5SallyChangeStratVecE2-2}]{\includegraphics[width=.45\linewidth]{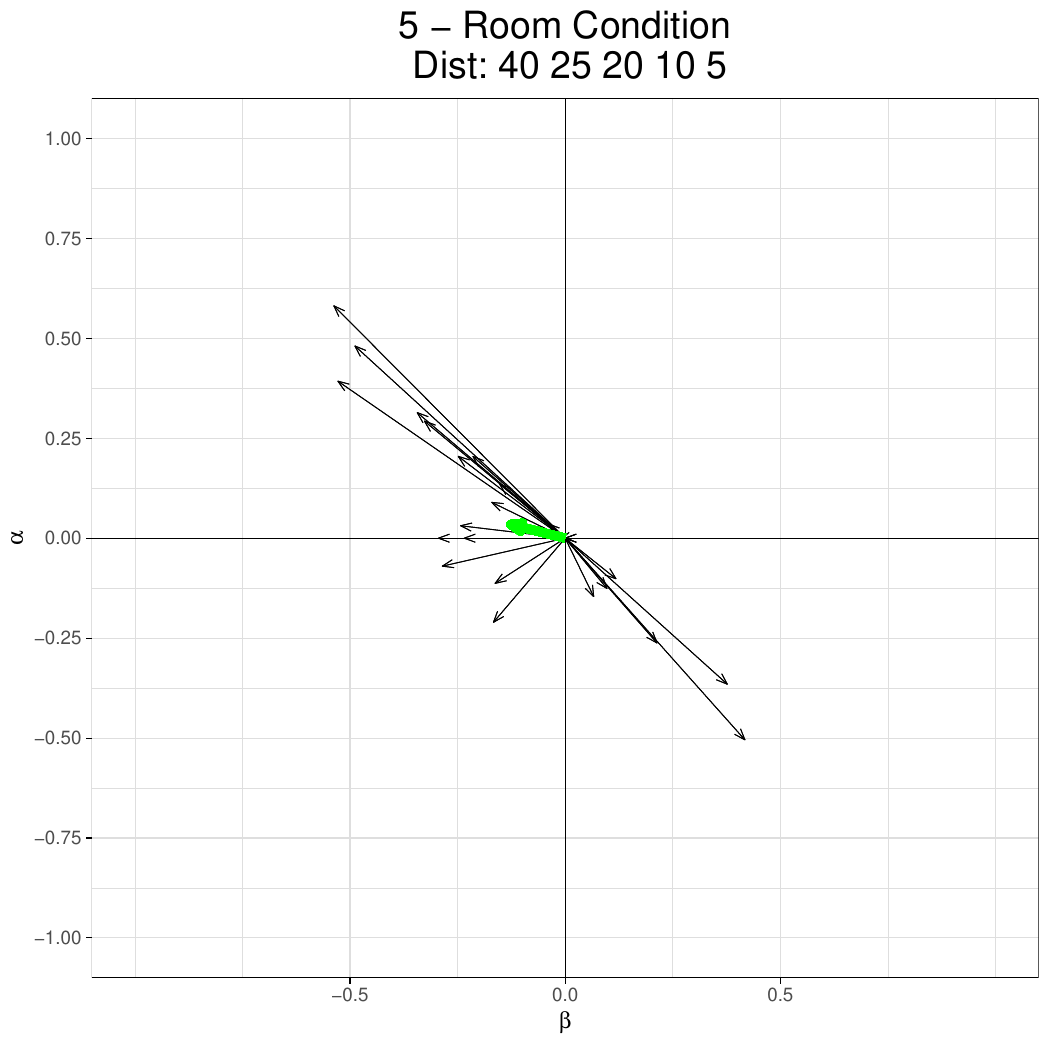} }\caption[Plotting Changes in Strategy Vectors]{Plotting Changes in Strategy Vectors: Two and Five Room Conditions Sally Distribution. The left plot shows the two-room condition and the right plot shows the five-room condition. For each plot the choice data for each participant was fit to best explain their choice data using a basis vector of the maximizing/minimizing and matching/antimatching strategies. These two vectors are subtracted to yield a change vector. This is plotted and the average of all the individual change vectors (thin black lines) is shown as a thick green line.}\label{fig:s2-5SallyChangeStratVecE2}
\end{figure}

\end{knitrout}

Unlike in Experiment 1, we did not see a notable difference in the strategy mix between the two-room and the five-room conditions. This appears to be largely driven by the fact that participants in Experiment 2 did not probability match as much as participants in Experiment 1, but instead used more maximizing, so we still observe that participant strategies lie close to the line $\alpha + \beta = 1$ implying our model appropriately accounts for participant hiding and seeking behavior in both two-room and five-room conditions (Figure \ref{fig:osally2seekStrat}). Again, the line $\alpha + \beta = 1$ represents a strategy mix of exclusively matching and maximizing for seeking, or exclusively antimatching and minimizing for hiding. If there is any consistent migration between the two conditions it is generally in favor of more minimization when hiding, and there is a numerically greater trend in the five room condition.

\subsection{Special case opponent strategies}

Participant behavior against Kala (uniform) and Bo (100\%) followed closely that of Experiment 1. While all strategies against a uniform distribution are equivalent, participants mostly produce a nearly uniform distribution, while hiding and seeking. Note that the reflection of the uniform is the uniform, thus the vector reflection hypothesis appears to still function in this ambiguous case.

Participant seeking against the 100\% distribution utilized the unique reasonable strategy -- only picking the 100\% room. With respect to hiding, participants show a mix between exploiting a particular 0\% room, and distributing hides across all of the 0\% rooms, interesting since all of which are equally good strategies in our task environment. 

The first two experiments show broadly consistent results. Participants are sensitive to the specific probability distributions used by their simulated opponents for hiding and seeking. They modulate their use of the probability information depending on the context they find themselves in: hiding or seeking. A combination of matching/antimatching and maximizing/minimizing strategies provides a good account of participant choices even in the five room situation when there is no mathematical necessity that it should. Additionally, the combination of matching/antimatching and maximizing/minimizing strategies participants deploy is modulated by the complexity of the distribution, in this case indexed by the number of rooms. Other distributions, like the uniform or one 100\% room, show that our participants adapt sensibly. When there is no benefit to adjusting a strategy combination for either gaining information or improving outcomes (uniform) the participants mathematically remain stable in their choices across contexts. When there is little need to track opponent choices, and little ambiguity about consequences, participants make the optimal choice. Against the always-choose-the-same-room opponent, matching and maximizing are basically the same thing. These distributions inspect the edge-cases of our model, and at the very least provide confirmation that participants are sensitive to the task manipulations and behave reasonably across various distributions, including these ``degenerate'' ones.

\section{Results Experiment 3}

\begin{figure}[tbp]
\begin{tikzpicture}
\begin{axis}[clip=false,axis lines=middle,
    axis equal image,
    xmin=0,
    xmax=1.6,
    ymin=0,
    ymax=1.15,
    zmin=0,
    zmax=1.15,
    xtick={1},
    ytick={1},
    ztick={1},
    view={80}{15}]
  \draw[line width=1.5pt, -stealth] (0,0,0)--(0.0,1/6,5/6) node[anchor = west]{$\vec{p}$};
  \draw[line width=1.5pt, -stealth] (0,0,0)--(2/3, 1/2, -1/6) node[anchor = west]{$\vec{q}$};
  \draw[line width=1.5pt, -stealth] (0,0,0)--(1/3, 1/3, 1/3) node[anchor = west]{$\vec{u}$};
  \draw[line width=1.5pt, red, -stealth] (2/3, 1/2, -1/6)--(0.5555556, 0.4444444, 0.0000000) node[anchor = west]{$\vec{-v}$};
\addplot3[patch, patch type=triangle, color=blue!25, fill opacity=0.6, 
faceted color=black, line width=0.5pt] 
coordinates{(1,0,0) (0,1,0) (0,0,1)};
\draw[line width=1pt, loosely dotted,-stealth](0.0,1/6,5/6)--(2/3, 1/2, -1/6);
\end{axis}
\end{tikzpicture}
\begin{tikzpicture}
  \begin{axis}[clip=false,width=0.7*\textwidth,
    axis lines=middle,
    axis equal image,
    xmin=0,
    xmax=1.25,
    ymin=0,
    ymax=1.25,
    zmin=0,
    zmax=1.25,
    xtick={1},
    ytick={1},
    ztick={1},
    view={135}{30}]
  \draw[line width=1.5pt, -stealth] (0,0,0)--(0.0,1/6,5/6) node[anchor = west]{$\vec{p}$};
  \draw[line width=1.5pt, -stealth] (0,0,0)--(2/3, 1/2, -1/6) node[anchor = west]{$\vec{q}$};
  \draw[line width=1.5pt, red, -stealth] (2/3, 1/2, -1/6)--(0.5555556, 0.4444444, 0.0000000) node[anchor = west]{$\vec{-v}$};
  \addplot3[patch, patch type=triangle, color=blue!25, fill opacity=0.6, 
    faceted color=black, line width=0.5pt] 
  coordinates{(1,0,0) (0,1,0) (0,0,1)};
\end{axis}
\end{tikzpicture}
\begin{tikzpicture}
  \begin{axis}[clip=false,width=0.7*\textwidth,
    axis lines=middle,
    axis equal image,
    xmin=0,
    xmax=1.25,
    ymin=0,
    ymax=1.25,
    zmin=0,
    zmax=1.25,
    xtick={1},
    ytick={1},
    ztick={1},
    view={135}{30}]
  \draw[line width=1.5pt, -stealth] (0,0,0)--(0.0,1/6,5/6) node[anchor = west]{$\vec{p}$};
  \draw[line width=1.5pt, -stealth] (0,0,0)--(2/3, 1/2, -1/6) node[anchor = west]{$\vec{q}$};
  \draw[line width=1.5pt, red, -stealth] (2/3, 1/2, -1/6)--(0.5833333, 0.4166667, 0.0000000) node[anchor = west]{$\vec{-v}$};
  \addplot3[patch, patch type=triangle, color=blue!25, fill opacity=0.6, 
    faceted color=black, line width=0.5pt] 
  coordinates{(1,0,0) (0,1,0) (0,0,1)};
\end{axis}
\end{tikzpicture}%
\caption{Three different viewpoints of the probability simplex in 3-dimensional space. Here, the reflection of $\vec{p}$, denoted $\vec{q}$, lies outside of the probability simplex. Performing the Euclidean projection onto the simplex (right most panel) produces the shortest possible vector $\vec{v}$ to the simplex, but does not always shift $\vec{q}$ back along the original reflection trajectory from $\vec{p}$ (as seen in the middle panel).}
\label{fig:outside}
\end{figure}

Experiment 3 increased the maximum number of rooms participants faced. This was to help distinguish if scenario complexity would influence how participants combined matching/antimatching and maximizing/minimizing strategies. Additionally, we included a wider range of distributions so as to have some distributions that had reflections outside of the probability simplex. This was to probe our suggestion that the cognitive operation for calculating an opposite distribution when someone is switching between the roles of hider and seeker (or for pursuit versus avoidance) is akin to vector reflection.

\subsection{Vector projection methods account for invalid reflections}

We use vector reflection of the matching strategy to create the antimatching strategy. We need to select a particular procedure, because there is no mathematically well-defined ``opposite'' distribution. The choice of vector reflection capitalizes on our geometric formalization to derive a second distribution from the matching distribution that is equidistant from the uniform, ala Bayes, and opposite in the geometric sense of being a reflection. And this operationalization of the notion of ``opposite'' brings with it the opportunity to test the vector reflection approach, because while it is always possible to reflect a vector across the uniform, it is not always possible to keep it in the positive quadrant. Thus, if people are really doing some operation akin to vector reflection they will have to deal with reflections that result in vectors that do not form permissible choice distributions (this is because some of the elements will be negative and you cannot make a choice a negative number of times). To illustrate the two methods that we test and to see an illustration of the program consult Figure~\ref{fig:outside}. One way to handle an invalid reflection is to return along the projection trajectory until you reach the simplex (or equivalently halt the reflection when you hit the probability simplex ``wall''). A second method is to project back to the simplex at its closest point. It is likely that any reasonable method to project an invalid reflection back to the simplex will result in very similar distributions. Future mathematical work is needed to determine which methods may differ the most, and which particular reflections may result in the largest discrepancy between projection methods. However, we do find that both projections produce dramatic improvements in our ability to model participant choice counts (for example see specifically the seven room condition in Figure~\ref{fig:d7c_projection_plot}). This implies that when a reflection does not exist within the simplex, and therefore the strategy cannot be expressed behvaiorally (since it has negative values), pariticpants instead use a distribution within the simplex that is near the invalid reflection. 

\begin{knitrout}
\definecolor{shadecolor}{rgb}{0.969, 0.969, 0.969}\color{fgcolor}\begin{figure}
\includegraphics[width=\maxwidth]{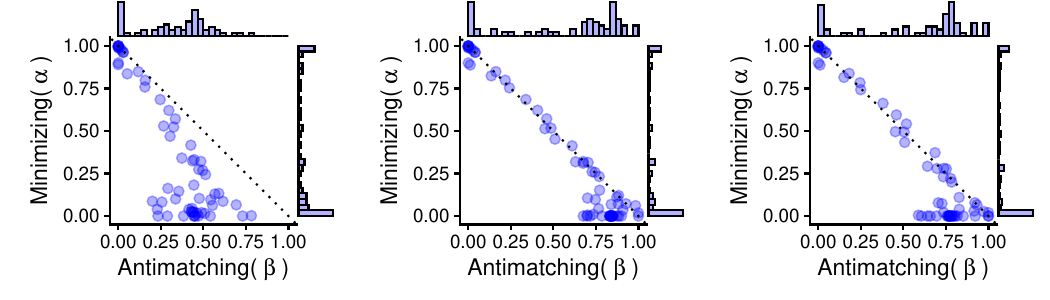} \caption[Defining pariticipant hide strategies using the original invalid reflection (left), unishift projection (center), and closest simplex point projection (right) in the 7 Room (50,18,12,8,5,5,2) condition]{Defining pariticipant hide strategies using the original invalid reflection (left), unishift projection (center), and closest simplex point projection (right) in the 7 Room (50,18,12,8,5,5,2) condition. Both methods of projecting back to the simplex, either along the path of reflection or to the nearest point, result in considerable improvement in the fit to participants' choice probabilities.}\label{fig:d7c_projection_plot}
\end{figure}

\end{knitrout}

\subsection{Strategy mixes change between hiding and seeking and across room-count conditions}

Figure \ref{fig:e3ChangeStratVec} provides an overview of how strategy mixes change between seeking and hiding. Comparing participant changes in strategy combinations between hiding and seeking conditions reveals that, in general, participants use the computationally simpler, yet optimal, minimizing strategy more when hiding than they use the analagous maximizing strategy while seeking. However, the strength of this effect does not obviously correlate to room-count as we might expect if computational complexity were a primary driver of policy mixing. Moreover, Figure~\ref{fig:allRoomHideStratChangeBox} indicates that as situational complexity (room-count) increases, participants use the mimimizing strategy when hiding \emph{less} in aggregate. 

\clearpage
\blandscape
\begin{knitrout}
\definecolor{shadecolor}{rgb}{0.969, 0.969, 0.969}\color{fgcolor}\begin{figure}[h]
\includegraphics[width=\maxwidth]{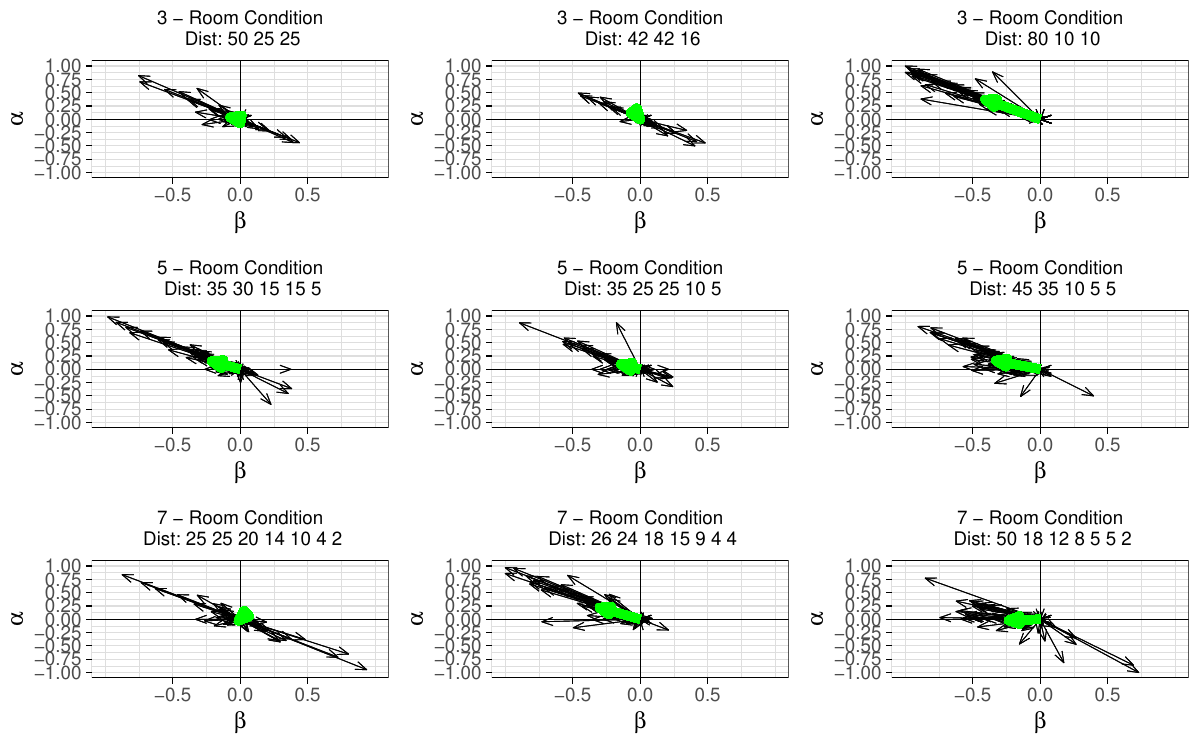} \caption[Strategy Changes for Different Numbers of Rooms]{Strategy Changes for Different Numbers of Rooms. In general, there is more use of the optimal strategy when hiding than when seeking (green arrows point up and to the left). The increase in the number of rooms increases the complexity of the task. Participants shift their strategy mix based on decision context.}\label{fig:e3ChangeStratVec}
\end{figure}

\end{knitrout}
\elandscape

\begin{knitrout}
\definecolor{shadecolor}{rgb}{0.969, 0.969, 0.969}\color{fgcolor}\begin{figure}
\includegraphics[width=\maxwidth]{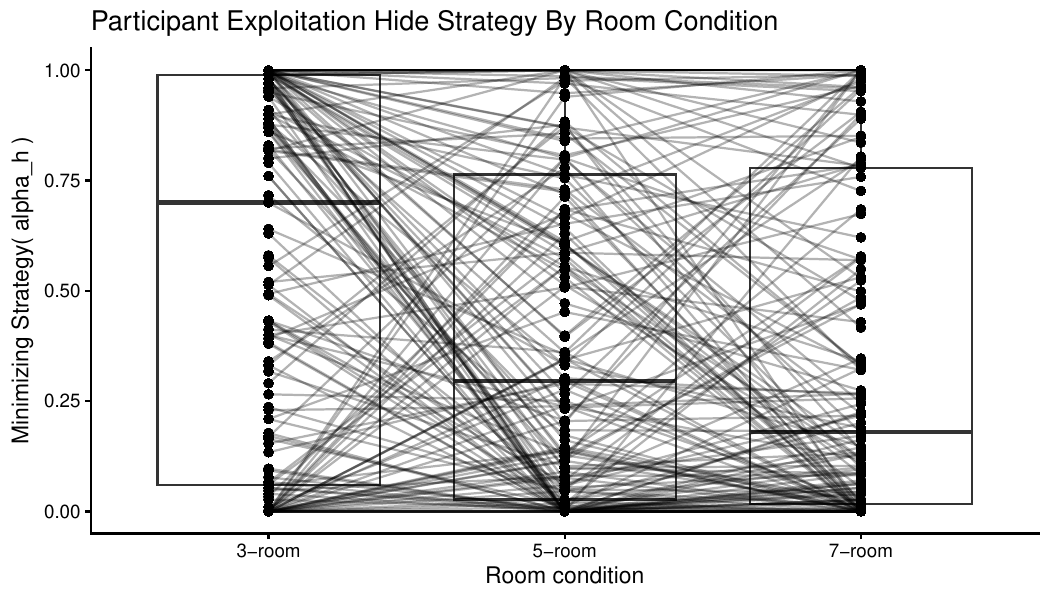} \caption[Participant use of the minimizing hiding strategy decreases as dimensionality increases]{Participant use of the minimizing hiding strategy decreases as dimensionality increases. Boxplots the minimizing strategy component by room-number condition, and also plots the individual participant strategy mix for each condition. Lines connect individual participant strategies across conditions.}\label{fig:allRoomHideStratChangeBox}
\end{figure}

\end{knitrout}

Taken together, we conclude that participants have a higher propensity to use a matching/antimatching strategy as situational complexity increases. This might be because this strategy provides a greater information gain, and thus might be viewed as analogous to exploration. We also find that for any given level of complexity, participants employed an optimal strategy more when hiding than when seeking. To explain this, one could consider an \emph{assumed-payoff} account, where hiding strategies are generally more exploitative than when seeking because participants hold prior beliefs that a failed hide is worse than a failed seek. Perhaps with more room options, the perceived importance of each individual hiding trial outcome is lesser because the overall probability of being found is lesser, thus reducing motivation to select a strict minimizing strategy when hiding. This account can be tested by manipulating payoffs of successful/failed hide/seeks via gamified points, money, and the like.

It is likely that computational complexity and assumed-payoffs jointly influence participants' behaviour. Even so, it is still true that n-dimensional geometry is harder to work with than m-dimensional geometry where n $>$ m. This increases the computational cost of determining the opposite probability distribution as room-count increases, thereby disincentivizing an antimatching strategy when hiding. The reduction in risk of being found when hiding as room-count increases may simultaneously \emph{incentivize} antimatching as a form of exploration, thus moderating the overall effect.

While we may not be able to tell which projection correction method is better with these data we note that each projection method comes paired with distinct theoretical implications. For the stop-at-the-edge account there need not be an internal representation of the invalid vector reflection. This would imply humans only represent what exists within the probability simplex in order to compute opposite probability distributions. Alternatively, the projection back to the nearest point on the simplex implies that humans can represent a broader interpretation of probabilities including those that are negative as long as they sum to 1. Under this account, probability could be represented like any other geometric/visuospatial representation. The closest point on the simplex would then only be needed to permit behavioral expression of the representation.

The benefit of our analysis here is that future researchers can propose new methods to account for invalid reflections. The effectiveness of alternative models can be directly compared to each other by examining the quantitative differences between strategy vector coefficients, and model error. 

\section{General Discussion}

We found that modeling people's choices as vectors provided a concise summary of choice behavior with a geometric flavor.  Whether seeking or hiding, regardless of room number, or opponent probability distribution, a linear combination of two canonical strategies maximizing/minimizing with matching/antimatching did a good job of accommodating participant choice frequencies. All four strategies can be computed from the matching strategy, which itself ony requires counting and remembering. Our most novel proposal is that of probability antimatching as a behavioral pattern in choice tasks emphasizing avoidance, and which we formalized as a Euclidean vector reflection. 

Experiment 1 established the efficacy of the hide-and-seek task in measuring pursuit and avoidance strategies. Experiment 2  was a replication of Experiment 1 using online materials. While there were some differences in the details, we found similar results. The data from these two studies showed the effectiveness of histogram methods for describing human choice policies, but it did not test the consistency of this two strategy combination account with varying complexity nor the mathematical boundary of the vector reflection account. Experiment 3 addressed this deficit. We adopted a within-subjects design so that every participant faced at least one 3-room, 5-room, and 7-room  condition. In addition, some participants also faced distributions that had invalid reflections. Again, we found a near exclusive mix of maximizing and matching when seeking, and a near exclusive mix of minimizing and antimatching when hiding. We also found that when hiding against distributions with invalid reflections, participants' antimatching strategy selected a distribution on the probability simplex that was ``close'' to the actual, but invalid, reflection. It is unclear exactly how the ``close'' distribution was derived. We considered two methods for this: backtracking along the line of reflection to the first ``legal'' histogram, and projecting from the point of reflection to the nearest point on the simplex. For our design, both selection methods led to very similar predictions. Finding a better way to separate these two accounts is a goal for future work, but the fact that two different geometrical accounts lead to, in practice, similar predictions means that the practical consequences for using either method is small.  

In both Experiments 1 and 3 (but less so in Experiment 2) participants utilized a more optimal strategy mix when hiding than when seeking. We interpret this as an asymmetric weighting of gains and losses and consistent with prospect theory \cite{kahneman1979prospect}. It has also been suggested from choice behavior data that severe risk may influence estimates of probabilities \cite{Weber1994}, but our data and analyses suggest that frequencies can be estimated separately from outcome severity. We suggest that what happens when severity and risk intrude is that participants will mix their two basis policies, matching/antimatching and maximizing/minimizing, differently. 

One of our motivations for proposing our frequency account is the fact that many real world processes are continuous and thus too complex to expect any precise estimates by human beings. We wanted to see how well frequency estimates worked and how they scaled when task complexity varied. We operationalized increasing complexity as increasing the number of rooms. Given how well a discrete model like ours works, we conjecture that when people are confronted by continuous real world problems there may be a feasibility constraint that leads to the adoption of a simpler, discrete model like the one we propose here. 

Our work demonstrates that participant choice frequencies for seeking are adequately defined as a mix of two strategies: traditional probability matching and optimal maximizing. We describe a novel strategy ``antimatching'' revealed by hiding behavior. We articulate human stimulus avoidance strategies with just two dimensions: antimatching and minimizing. Antimatching and minimizing are direct analogues of matching and maximizing.  It is noteworthy that we can achieve such predictive success despite there being no established unambiguous mathematical definition of an opposite probability distribution. We highlight that the distinction between the hiding/seeking conditions are more a reflection of task demands and the participant's perspective than they are differences in a fundamental representation of probability. 

Across our three studies, we observe a pattern of findings where participants use a more optimal strategy when hiding than they do when seeking. The strength of this finding, and the manner in which it is affected by the number of room choices (dimensionality) is unclear, and will require further study. Our current explanation for this shift towards optimality is an intuition from evolutionary considerations: a failed hide often carries greater consequence (eaten by a predator) than a failed seek (being hungry). This might induce a default payoff matrix that skews to value the outcomes of hide success over seek success. By modifying our hide-and-seek task to accommodate different payoffs, via arbitrary in-task points, monetary, or course credit it should be possible to test this intuition. And there may be practical benefits to confirming this idea. Many problems of daily life can be presented in a pursuit/avoidance dichotomy. It may be that simply framing a problem as something to hide from, rather than something to pursue, will make people act more optimally, at least in the short term. For example, would people make better financial decisions if the problem were framed as an avoidance of losses/costs (including opportunity cost) rather than a pursuit of gains \cite{soman2004framing, levin1998all, kahneman1979prospect}. 

Our work presents a testable hypothesis for strategy utilization that accounts for both stimulus pursuit and stimulus evasion, therefore creating a meaningful extension to current probability matching literature. The formal geometric terms used to represent participant strategy mixes allow for richer analysis and quantitative theoretical development. For example, if another researcher proposes an alternative definition of antimatching, we can again represent participant strategy mix as a linear combination of Euclidean vectors and compare model fit via the magnitudes of strategy coefficients and model error.

With this work, we are not advocating that people are literally and explicitly computing vector reflections to determine their hiding strategies. Claiming this would imply that people are performing non-trivial algebra in their heads in order to complete our task. The benefit of a model with geometric interpretations is that it can be explained and conceptualized \emph{without} algebra. This is similar to how catching a thrown ball is fundamentally different than solving a physics problem about a thrown ball. Our model with a built-in geometric interpretation, holds a ``feasibility advantage'' over other models such that the cognitive mechanisms required to emulate its output may already exist and be used for other visual-spatial capacities. Consider the analogy of Bayesian cognition. That is, the claim that people are ``Bayesian'' is infeasible as a literal claim since we cannot assume that people calculate the computationally difficult integrals in their head required by Bayes' rule, despite often presenting behavioral patterns that loosely emulate Bayesian models \cite{jones2011bayesian}. It may be that some other mechanism is the driving force of this emulating antimatching behavior. Claiming people to be literally Bayesian is equivalent to claiming that people literally ``use'' vector reflections. While the geometric nature of vector representations provides some degree of feasibility over and above literal Bayesian cognition, we still stress that our model is effective in describing \emph{behavior}. Cognitive and neural extensions of its implementation are left for future work.

\section{Acknowledgements}
We appreciate the contributions of Vivian DiBerardino for assistance with code editing and writing the code for strategy change plots. Britt Anderson was supported by an NSERC Discovery Award. The manuscript files, data analysis and plotting code, and the data are freely available at \url{https://codeberg.org/brittAnderson/prob-antimatch} .

\bibliography{ms}

\begin{appendix}
\section{Additional Distributions}
\centering  
\begin{knitrout}
\definecolor{shadecolor}{rgb}{0.969, 0.969, 0.969}\color{fgcolor}\begin{figure}
\includegraphics[width=\maxwidth]{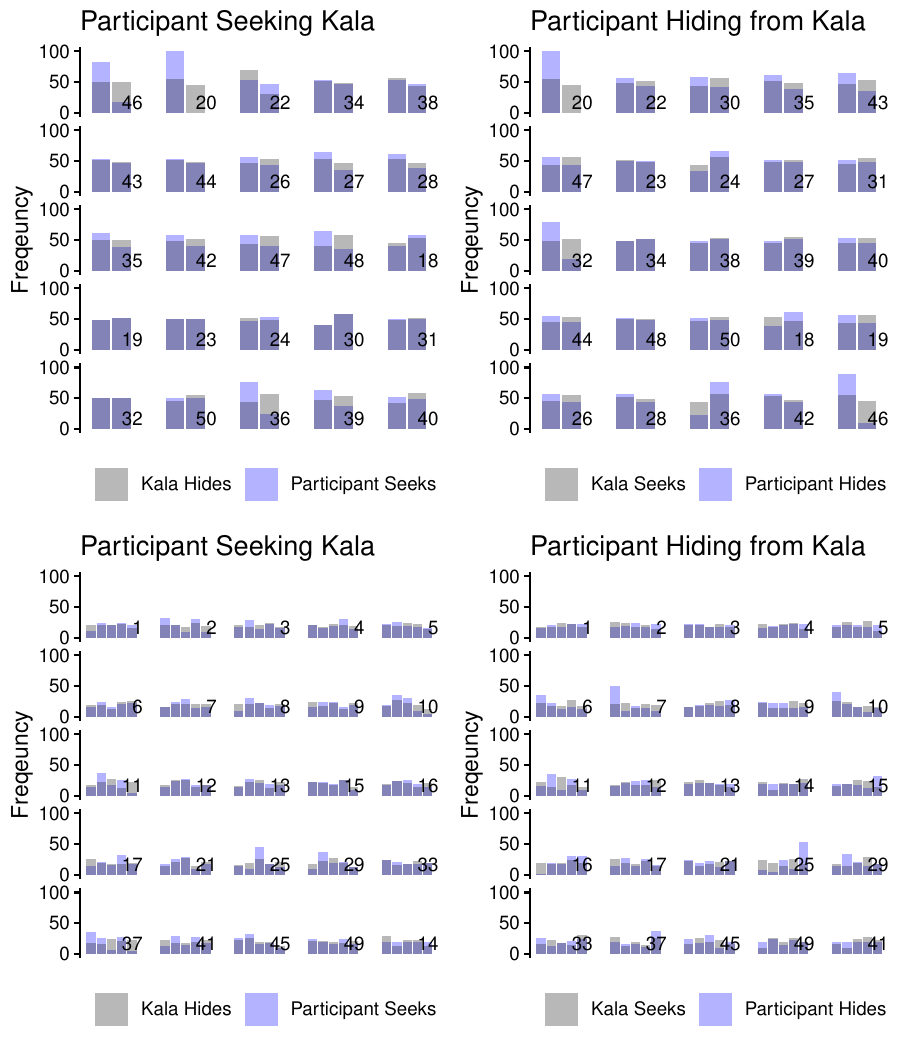} \caption[Histograms for Alternate Distributions (Kala)]{Histograms for Alternate Distributions (Kala). Confirming that participants understood the task and behaved accordingly we find that for an opponent distribution that chooses all options equally often ('Kala'; Two Room Condition Row 1 and Five Room Condition Row 2) the choice of the participant has no effect on their probability of winning and their default choice histogram largely resembles matching with some individual heterogeneity.}\label{fig:e1HistsAltDistKala}
\end{figure}

\end{knitrout}

\begin{knitrout}
\definecolor{shadecolor}{rgb}{0.969, 0.969, 0.969}\color{fgcolor}\begin{figure}
\includegraphics[width=\maxwidth]{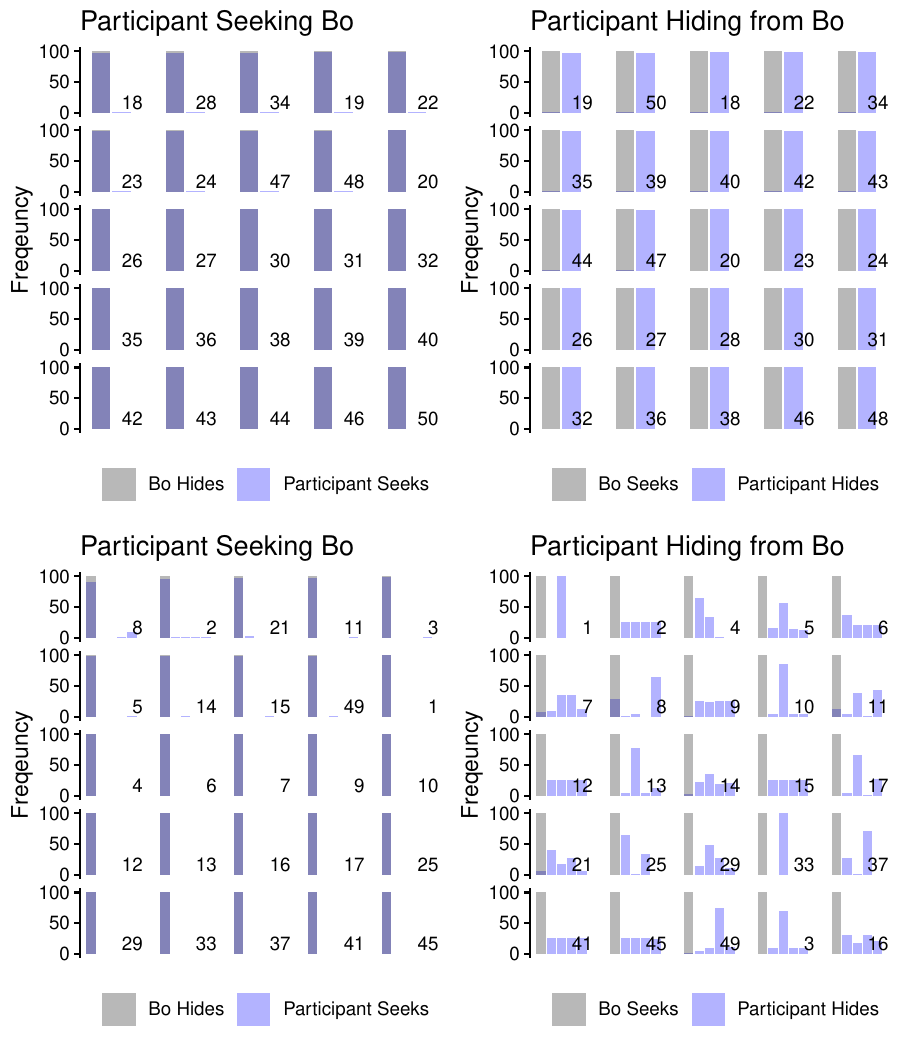} \caption[Histograms for Alternate Distributions]{Histograms for Alternate Distributions. Confirming that participants understood the task and behaved accordingly we find that for an opponent distribution that chooses one option 100\% of the time ('Bo'; Two Room Condition Row 1 and Five Room Condition Row 2) the participants always look there and never hide there.}\label{fig:e1HistsAltDistBo}
\end{figure}

\end{knitrout}
\end{appendix}
\end{document}